\documentclass[12pt,preprint]{aastex}
 
\newcommand{\etal}{et~al.\ } 
\newcommand{\PVdblt}{{\rm P}\kern 0.1em{\sc v}~$\lambda\lambda 1117, 1128$} 
\newcommand{\CaIIdblt}{{\rm Ca}\kern 0.1em{\sc ii}~$\lambda\lambda 3934, 3969$}
\newcommand{\AlIIIdblt}{{\rm Al}\kern 0.1em{\sc iii}~$\lambda\lambda 1854, 1862$} 
\newcommand{\CIVdblt}{{\rm C}\kern 0.1em{\sc iv}~$\lambda\lambda 1548, 1551$} 
\newcommand{\MgIIdblt}{{\rm Mg}\kern 0.1em{\sc ii}~$\lambda\lambda 2796, 2803$} 
\newcommand{\NVdblt}{{\rm N}\kern 0.1em{\sc v}~$\lambda\lambda 1239, 1243$}
\newcommand{\SVIdblt}{{\rm S}\kern 0.1em{\sc vi}~$\lambda\lambda 933, 944$} 
\newcommand{\OVIdblt}{{\rm O}\kern 0.1em{\sc vi}~$\lambda\lambda 1032, 1038$} 
\newcommand{\SiIIdblt}{{\rm Si}\kern 0.1em{\sc ii}~$\lambda\lambda 1190, 1193$} 
\newcommand{\SiIVdblt}{{\rm Si}\kern 0.1em{\sc iv}~$\lambda\lambda 1394, 1403$} 
\newcommand{\PV}{\hbox{{\rm P}\kern 0.1em{\sc v}}} 
\newcommand{\AlI}{\hbox{{\rm Al}\kern 0.1em{\sc i}}} 
\newcommand{\AlII}{\hbox{{\rm Al}\kern 0.1em{\sc ii}}}
\newcommand{\AlIII}{{\hbox{\rm Al}\kern 0.1em{\sc iii}}}
\newcommand{\CaII}{\hbox{{\rm Ca}\kern 0.1em{\sc ii}}}
\newcommand{\CII}{\hbox{{\rm C}\kern 0.1em{\sc ii}}}
\newcommand{\CIIe}{\hbox{{\rm C$^{\ast}$}\kern 0.1em{\sc ii}}}
\newcommand{\CIII}{\hbox{{\rm C}\kern 0.1em{\sc iii}}}
\newcommand{\CIV}{\hbox{{\rm C}\kern 0.1em{\sc iv}}}
\newcommand{\CV}{\hbox{{\rm C}\kern 0.1em{\sc v}}}
\newcommand{\HI}{\hbox{{\rm H}\kern 0.1em{\sc i}}}
\newcommand{\HII}{\hbox{{\rm H}\kern 0.1em{\sc ii}}}
\newcommand{\Lya}{\hbox{{\rm Ly}\kern 0.1em$\alpha$}}
\newcommand{\Lyb}{\hbox{{\rm Ly}\kern 0.1em$\beta$}}
\newcommand{\Lyg}{\hbox{{\rm Ly}\kern 0.1em$\gamma$}}
\newcommand{\Lyd}{\hbox{{\rm Ly}\kern 0.1em$\delta$}}
\newcommand{\Lye}{\hbox{{\rm Ly}\kern 0.1em$\epsilon$}}
\newcommand{\Lyphi}{\hbox{{\rm Ly}\kern 0.1em$\phi$}}
\newcommand{\Lyfive}{\hbox{{\rm Ly}\kern 0.1em$5$}}
\newcommand{\Lysix}{\hbox{{\rm Ly}\kern 0.1em$6$}}
\newcommand{\Lyseven}{\hbox{{\rm Ly}\kern 0.1em$7$}}
\newcommand{\Lyeight}{\hbox{{\rm Ly}\kern 0.1em$8$}}
\newcommand{\Lynine}{\hbox{{\rm Ly}\kern 0.1em$9$}}
\newcommand{\Lyten}{\hbox{{\rm Ly}\kern 0.1em$10$}}
\newcommand{\Lyeleven}{\hbox{{\rm Ly}\kern 0.1em$11$}}
\newcommand{\HeI}{\hbox{{\rm He}\kern 0.1em{\sc i}}}
\newcommand{\HeII}{\hbox{{\rm He}\kern 0.1em{\sc ii}}}
\newcommand{\FeI}{\hbox{{\rm Fe}\kern 0.1em{\sc i}}}
\newcommand{\FeII}{\hbox{{\rm Fe}\kern 0.1em{\sc ii}}}
\newcommand{\FeIII}{\hbox{{\rm Fe}\kern 0.1em{\sc iii}}}
\newcommand{\MnII}{\hbox{{\rm Mn}\kern 0.1em{\sc ii}}}
\newcommand{\MgI}{\hbox{{\rm Mg}\kern 0.1em{\sc i}}}
\newcommand{\MgII}{\hbox{{\rm Mg}\kern 0.1em{\sc ii}}}
\newcommand{\MgIII}{\hbox{{\rm Mg}\kern 0.1em{\sc iii}}}
\newcommand{\NI}{\hbox{{\rm N}\kern 0.1em{\sc i}}}
\newcommand{\NII}{\hbox{{\rm N}\kern 0.1em{\sc ii}}}
\newcommand{\NIII}{\hbox{{\rm N}\kern 0.1em{\sc iii}}}
\newcommand{\NV}{\hbox{{\rm N}\kern 0.1em{\sc v}}}
\newcommand{\OVI}{\hbox{{\rm O}\kern 0.1em{\sc vi}}}
\newcommand{\OI}{\hbox{{\rm O}\kern 0.1em{\sc i}}}
\newcommand{\OII}{\hbox{[{\rm O}\kern 0.1em{\sc ii}]}}
\newcommand{\OIV}{\hbox{{\rm O}\kern 0.1em{\sc iv}]}}
\newcommand{\SI}{{\rm S}\kern 0.1em{\sc i}} 
\newcommand{\SIV}{{\rm S}\kern 0.1em{\sc iv}} 
\newcommand{\SVI}{{\rm S}\kern 0.1em{\sc vi}}
\newcommand{\SiI}{\hbox{{\rm Si}\kern 0.1em{\sc i}}}
\newcommand{\SiII}{\hbox{{\rm Si}\kern 0.1em{\sc ii}}}
\newcommand{\SiIII}{\hbox{{\rm Si}\kern 0.1em{\sc iii}}}
\newcommand{\SiIV}{\hbox{{\rm Si}\kern 0.1em{\sc iv}}}
\newcommand{\SII}{\hbox{{\rm S}\kern 0.1em{\sc ii}}}
\newcommand{\SIII}{\hbox{{\rm S}\kern 0.1em{\sc iii}}}
\newcommand{\NaI}{\hbox{{\rm Na}\kern 0.1em{\sc i}}}
\newcommand{\TiII}{\hbox{{\rm Ti}\kern 0.1em{\sc ii}}}
\newcommand{\kms}{\hbox{km~s$^{-1}$}}
\newcommand{\cmsq}{\hbox{cm$^{-2}$}}
\newcommand{\cc}{\hbox{cm$^{-3}$}}

\tighten
\begin{document}
 
 


\title{The Multi--phase Absorption Systems Toward PG 1206+459\altaffilmark{1,2,3}}

\author{Jie~Ding, Jane~C.~Charlton\altaffilmark{4},
Christopher~W.~Churchill\altaffilmark{5,6}, and Christopher Palma}
\affil{Department of Astronomy and Astrophysics \\ The Pennsylvania
State University \\ University Park, PA 16802 \\ {\it ding, charlton,
cwc, cpalma@astro.psu.edu}}

\affil{The Pennsylvania State University, University Park, PA 16802}

\altaffiltext{1}{Based in part on observations obtained at the
W.~M. Keck Observatory, which is operated as a scientific partnership
among Caltech, the University of California, and NASA. The Observatory
was made possible by the generous financial support of the W. M. Keck
Foundation.}
\altaffiltext{2}{Based in part on observations obtained with the
NASA/ESA {\it Hubble Space Telescope}, which is operated by the STScI
for the Association of Universities for Research in Astronomy, Inc.,
under NASA contract NAS5--26555.}
\altaffiltext{3}{Based in part on observations obtained with the WIYN 
3.5--m telescope, a joint facility of the University of
Wisconsin--Madison, Indiana University, Yale University, and the
National Optical Astronomy Observatories.}
\altaffiltext{4}{Center for Gravitational Physics and Geometry}
\altaffiltext{5}{Visiting Astronomer at the W.~M. Keck Observatory}
\altaffiltext{6}{Visiting Astronomer, Kitt Peak National Observatory, 
National Optical Astronomy Observatory, which is operated by the
Association of Universities for Research in Astronomy, Inc. (AURA)
under cooperative agreement with the National Science Foundation.}

\begin{abstract}

A high--resolution ($R=30,000$) ultraviolet spectrum is presented,
which covers {\Lya} and many low--, intermediate--, and
high--ionization transitions in the three {\MgII}--selected absorption
systems toward the quasar PG~$1206+459$. Three systems (A, B, and C),
which are clustered within $1500$~{\kms} at $z \sim 0.93$, were
originally identified in a spectrum obtained with the High Resolution
Spectrograph (HIRES) on the Keck I telescope.  A WIYN Gunn $i$--band
image of the quasar field and spectroscopy of two galaxy candidates
are presented.  A multi--phase medium is seen in all three systems,
consistent with smaller, denser clouds producing low--ionization
transitions ({\MgII}, {\FeII}, and {\SiII}) and larger, diffuse,
photoionized clouds giving rise to higher--ionization transitions
({\CIV}, {\NV}, and/or {\OVI}). (1) System A, a multi--cloud, weak
{\MgII} absorber at $z=0.9254$, requires a super--solar metallicity in
both low-- and high--ionization phases, unless an $\alpha$--group
enhancement is included. The low--ionization absorption is produced in
clouds with sizes of $10$--$70$~pc, which are surrounded in velocity
space by broader, high--ionization components. With the unusually
complex velocity structure resolved in the {\NV} profiles, this system
is unlikely to represent a traditional galaxy disk/corona. The most
likely candidate host galaxy is a $\sim 2L^*$, apparently warped,
spiral at an impact parameter of $43 h^{-1}$~kpc. (2) System B, at
$z=0.9276$, has the strongest {\MgII} absorption and has an
approximately solar metallicity in the low--ionization phase. The
smooth, broad high--ionization profiles may indicate a coronal
structure similar to that of the Milky Way. The redshift of an $L^*$
galaxy ($z=0.9289$), at an impact parameter of $38 h^{-1}$~kpc is
consistent with the redshift of this system. (3) System C, at
$z=0.9342$, has a single component in {\MgII}, separated from the
other two systems by $\sim +1000$~{\kms}.  The {\Lya} profile is not
aligned with the {\MgII}, requiring an additional velocity component
offset by $-40$~{\kms}. System C lacks the small, low--ionization
cloud characteristic of an isolated single--cloud, weak {\MgII}
absorber.  Its absorption properties are similar to the ``satellite
clouds'' of classic strong {\MgII} absorbers, so this could be a
high--velocity cloud in the galaxy group responsible for the systems,
possibly related to a $0.2L^*$ galaxy at an impact parameter of $43
h^{-1}$~kpc.

\end{abstract}


\section{Introduction}
\label{sec:intro} 

Quasar absorption line systems, selected by {\MgII} absorption, sample
different  types   of  galaxies  with  absorption   produced  by  some
combination   of   disk   interstellar   medium,  coronal   gas,   and
high--velocity      clouds     \citep{disks,kinmod,steidel02}.     The
high--resolution spectra covering multiple chemical transitions reveal
detailed  information  of  the  kinematic,  chemical,  and  ionization
conditions in the different phases of  gas along the line of sight. At
$z \sim 1$, when the universe  was roughly half its present age, it is
of particular  interest to use  this information for  probing galactic
evolution.

Strong {\MgII} systems [$W_r (2796) > 0.3$~{\AA}] at $0.3 \la z \la 1$
are almost always found within $\sim 40 h^{-1}$~kpc of luminous
galaxies (those with luminosities greater than $0.05L^{\ast}$)
\citep{bb91,bergeron92,lebrun93,sdp94,s95,csv96,3c336}.  Therefore
they provide excellent probes of the gaseous content in galaxies at
intermediate redshifts.

Strong {\MgII} absorbers with twice the typical {\MgII}, {\FeII},
{\CIV}, and {\Lya} absorption have been classified as ``double''
systems \citep{archive2}.  These double systems also have twice the
typical kinematic spread in their low--ionization components. They
could be produced by lines of sight that pass through pairs of
galaxies \citep[see][]{archive2}, though at least one example exists
(the double system at $z = 0.8519$ toward Q~$0002+052$) where only one
galaxy has been identified as the absorbing galaxy \citep{csv96}.
This galaxy has a very blue rest--frame color and a {\it Hubble Space
Telescope\/} ({\it HST}) / Wide Field and Planetary Camera (WFPC2)
image reveals it to be morphologically compact and apparently isolated
\citep{steidel98}.  These properties are suggestive of an elevated
star formation rate.  Another interpretation, then, is that the double
systems could arise from material mechanically stirred up by star
formation processes \citep{archivelett}.

A particularly interesting example of the class of double, strong
{\MgII} absorbers is found at $z \sim 0.93$ along the line of sight
toward the quasar PG~$1206+459$ ($z_{em} = 1.16$).  Based upon spectra
obtained with the High Resolution Spectrograph (HIRES; \cite{vogt94})
on Keck~I, three separate ``groupings'' of low--ionization clouds were
identified \citep[labeled systems A, B, and C at redshifts $z=0.9254,
0.9276$, and $0.9343$, respectively;][]{q1206}.  Systems A and B were
together classified as a double {\MgII} system \citep{archive2},
whereas system C was classified as a ``weak system''
\citep{weak1,archive2}.  The low--resolution spectrum of
PG~$1206+459$, obtained with the {\it HST} / Faint Object Spectrograph
(FOS), is unusually rich, with strong absorption in the
high--ionization transitions \citep{kp13}.  \citet{q1206} studied the
{\it HST}/FOS and the HIRES/Keck spectra and, based upon
photoionization modeling of the data, inferred multiple ionization
phases in all three systems.  The low--ionization ``{\MgII} clouds''
appear to be embedded in extended, high--ionization, lower density gas
that gives rise to the bulk of the {\CIV}, {\NV}, and {\OVI}
absorption.

However, since in the previous study only low--resolution profiles
were available for many of the key transitions, the kinematic
relationships between the low-- and high--ionization gas remains
unknown.  The high--ionization profiles ({\CIV}, {\NV}, and {\OVI})
are resolved in the FOS spectrum (FWHM$\simeq 230$~{\kms}), indicating
a large velocity spread; however, it could not be determined whether
the profiles are smooth (suggesting a monolithic turbulent medium), or
are structured (suggesting smaller discrete high--ionization clouds).

As such,  a study of this  system holds the promise  of revealing
the comparative nature of the  low-- and high--ionization
gas in a  $z \simeq 1$ galaxy (or small group  of galaxies).  To better
understand the nature of this  system, we obtained an $R=30,000$ E230M
spectrum  from   the  Space  Telescope   Imaging  Spectrograph  (STIS)
on--board   {\it  HST}.    This  spectrum   provided  high--resolution
(FWHM$\simeq 15$~{\kms}) absorption  profiles for the {\Lya}, {\SiII},
{\CII}, {\SiIII}, {\SiIV}, {\CIV}  and {\NV} transitions documented in
the FOS spectrum.

\citet{q1206} discussed the possibility that the absorption may arise
in a small group of galaxies.  They tentatively reported a slight over
density of galaxies in a ground--based image \citep{kirhakos} of the
QSO field (they identified three bright galaxies within
5{\arcsec} of the QSO).  The extreme absorption in systems A + B,
their proximity in velocity, and the presence of a weak {\MgII} system
$\sim 1000$~{\kms} to the red, make the relationship between the
absorption and galaxies all the more tantalizing.  There are no other
clear galaxy candidates for {\MgII} absorption to arise in such a
potentially complex environment.  Do the absorption--galaxy
relationships suggest that the gaseous structures are coupled to
galaxies or dispersed throughout a small--group environment?

To address the absorption--galaxy relationship in this remarkable
multiple absorption line system, we obtained a WIYN 3.5--m telescope
Gunn $i$--band image of the QSO field (0.8{\arcsec} seeing).  We also
obtained spectra of the bright galaxies reported by \citet{kirhakos}
using the CryoCam Spectrograph on the 4--m telescope at Kitt Peak
National Observatory (KPNO).  We found that one of the galaxy
candidates identified by \citet{kirhakos} did not exist. We
successfully obtained the redshift of one of the other galaxies.

In this paper, we present the high--resolution profiles of the
absorption arising in systems A, B, and C.  We also present a WIYN
image of the QSO field.  Because not all useful transitions (i.e.\
higher order hydrogen lines and {\OVIdblt}) were observed at high
resolution, we also include the FOS spectrum.  We pursue
photoionization and collisional ionization modeling of the various
ionic transitions measured in both the STIS and HIRES spectra.  We
derive constraints on the metallicities, abundance patterns, and
ionization states of absorbing clouds in the different phases of gas.

In \S~\ref{sec:data} we briefly describe the Keck/HIRES, {\it
HST}/FOS, and {\it HST}/STIS spectra that we use to constrain our
models. Our WIYN image of the quasar field is presented in
\S~\ref{sec:galaid}, where we discuss our effort to spectroscopically
identify host galaxies of the three absorption systems.  A summary of
our photoionization/collisional ionization modeling techniques is
presented in \S~\ref{sec:methods} and the results of our modeling are
outlined in \S~\ref{sec:results}. Finally, in \S~\ref{sec:discussion},
we summarize and consider physical interpretations of the three
systems.


\section{The Data}
\label{sec:data}

\subsection{Keck/HIRES QSO Spectroscopy}
\label{sec:hires}

The optical  spectrum from HIRES  \citep{vogt94} extends from  3723 to
6186~{\AA},   and   covers    the   {\MgIIdblt}   doublet,   and   the
{\MgI}~$\lambda$2853,  {\FeII}~$\lambda$2344,  2374,  2383, 2587,  and
2600 transitions  at $z=0.93$.  The spectral  resolution is $R=45,000$
(FWHM $\sim 6.6$~{\kms}) and  the signal--to--noise ratio is $\sim 50$
per  three--pixel resolution  element  \citep{q1206,cv01}.  The  HIRES
spectrum was reduced in  the standard fashion using IRAF\footnote{IRAF
is distributed by the  National Optical Astronomy Observatories, which
are operated by AURA, Inc., under cooperative agreement with the NSF.},
as described  in \citet{cv01}.  The wavelength  calibration was vacuum
with a heliocentric correction applied.

\subsection{HST/STIS QSO Spectroscopy}
\label{sec:stis}

A  $24,000$  second  {\it  HST}/STIS  spectrum  of  PG~$1206+459$  was
obtained  with   the  E230M  grating   in  May  2001.    E230M  covers
$2270$~{\AA} to  $3120$~{\AA}.  The resulting  signal--to--noise ratio
ranges from  $5 \la$  S/N $\la 15$,  and the resolution  is $R=30,000$
(FWHM $\sim 10$~{\kms}). The total exposure time was $24,000$ seconds.
We reduced the STIS spectrum with the standard pipeline \citep{brown}.
The continuum  fitting was performed  on the extracted  spectrum using
standard methods  \citep{cv01}.  Absorption lines  were detected using
the formalism of \citet{schneider} and \citet{archive1}.

\subsection{HST/FOS QSO Spectroscopy}
\label{sec:fos}

Even with the availability of high--resolution STIS data, the FOS
spectrum is still critical in coverage of some key transitions that
are not within the STIS wavelength range. This includes {\OVIdblt} as
well as the Lyman series and a partial Lyman limit break at
$1760$~{\AA}.  The {\it HST}/FOS spectra \citep{kp13} covered
1600~{\AA} to 3280~{\AA} using the G190H and G270H gratings,
respectively, and had a resolution of $R=1,300$ (FWHM $\sim
230$~{\kms}). The data reduction, wavelength calibration, and
continuum fitting were performed as part of the QSO Absorption Line
Key Project \citep{kp1,kp7,kp13}.

\subsection{WIYN Imaging}

The field  surrounding PG~$1206+459$ was  observed on 9  February 1999
with the  WIYN 3.5--m  telescope.  The detector  (the 2048$\times$2048
CCD imager) provided a $6\farcm7  \times 6\farcm7$ field of view.  The
total  exposure time  of 5840  seconds  was divided  into 17  dithered
images  with integration  times of  300,  340, and  400 seconds.   All
observations were  taken through a  Gunn--$i$ filter.  The  seeing was
approximately  0.8\arcsec.  The  raw images  were processed  using the
IRAF package CCDRED.

Aperture photometry of several objects near the quasar was performed
using the IRAF APPHOT package.  An aperture of radius nine pixels
(approximately twice the FWHM of a star) was used to measure the
object flux; while the sky was determined by taking the centroid of
the pixel values inside an annulus with inner radius ten pixels and
outer radius 15 pixels.  It is likely that some light from the quasar
has affected the estimate of the sky flux, however, we calculated only
relative photometry for objects that are approximately equidistant
from the center of the QSO, therefore any systematic error in the sky
flux should not significantly affect our conclusions.

\subsection{CryoCam Galaxy Spectroscopy}

Spectra  of the two  galaxies G11  and G13
reported  by  \citet{kirhakos}  were  obtained with  the  Mayall  4--m
telescope at Kitt Peak National  Observatory on 15 February 1999.  The
CryoCam spectrograph  was used  with the 730  Grism, which  yielded an
approximate wavelength  coverage of $5500$--$10,000$~{\AA}.   The slit
was  oriented  at a  position  angle  of  $145${\arcdeg} in  order  to
simultaneously place both galaxies in the slit.

The  raw data  were processed  using the  IRAF CCDPROC  package.  Four
exposures   of  $2700$   seconds   each  were   obtained.   The   four
two--dimensional images  of the spectra  were compared, and  using the
positions of the night sky lines it was verified that the line centers
were  stable at  the sub--pixel  level.   Thus, we  combined the  four
exposures in the image plane using the IRAF IMCOMBINE task rather than
extracting  and combining the  one--dimensional spectra.   The spectra
were then extracted using the IRAF task APALL.

\section{Data Analysis}
\label{sec:dataanalysis}

\subsection{Absorption Line Systems}

\subsubsection{Optical Wavelengths}

The three absorption systems, designated A, B, and C, were originally
identified based upon the {\MgII} kinematics at redshifts $z=0.9254$,
$z=0.9276$, and $z=0.9342$ in the HIRES spectrum \citep{q1206}.  In
the top three panels of Figure~\ref{fig:data1}, we present the
absorption profiles in the rest--frame velocity of system B.  The
velocity zero point is defined at $z=0.927602$.  Twelve Voigt profile
(VP) components were used to model the {\MgII} profiles
\citep{q1206,cv01}.  The solid curve through the data are the model
spectra based upon AUTOVP MINFIT, where the numbered ticks above the
continuum provide the component numbers.  We will refer to these
numbers throughout our discussion in order to facilitate the identity
of individual ``{\MgII} clouds''.

In Table~\ref{tab:tab1}, we list, for each of the three systems, the
 rest--frame equivalent widths (and $3~\sigma$ limits for
 non--detections) of the {\MgIIdblt}, {\MgI}~2853, and
 {\FeII}~$\lambda 2600$ transitions.  The {\MgII} and {\FeII} column
 densities obtained from the VP decomposition were presented in
 \citet{q1206}.  The density of absorption lines associated with the
 three systems in the optical spectrum is relatively low, and at the
 resolution of HIRES, there are no blends between the systems nor
 interloping features from metal--line systems at other redshifts
 \citep[see][]{kp13}.

\subsubsection{Near Ultraviolet Wavelengths}
\label{sec:stisblends}

\begin{figure*}
\figurenum{1} 
\plotone{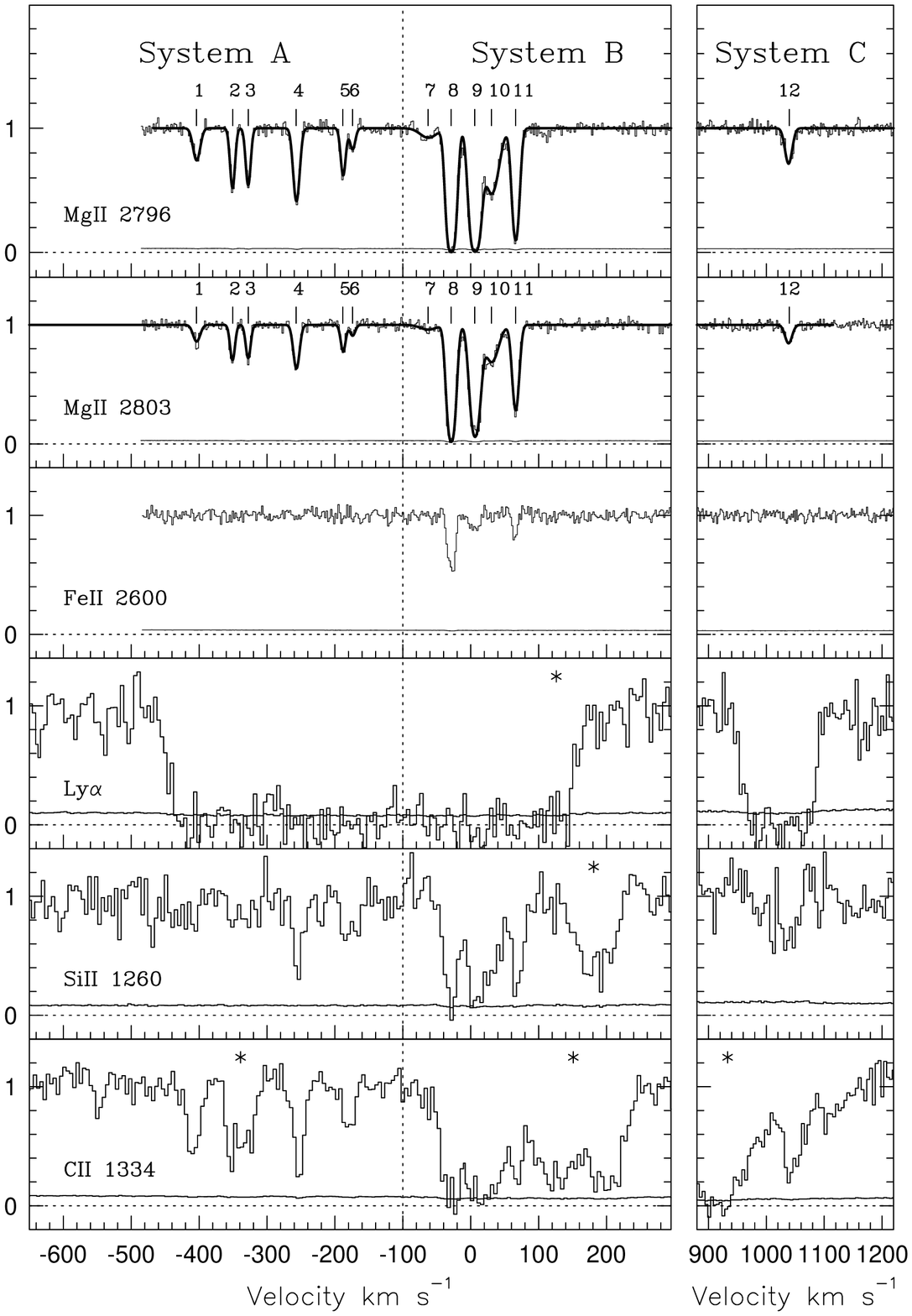} 
\vglue -0.8in 
\protect\caption 
{\footnotesize 
The spectra of various transitions are displayed as histograms in
velocity space in the figure above. The {\MgII} and {\FeII} profiles
are covered by HIRES/Keck ($R=45,000$), while the others are STIS/{\it HST}
($R=30,000$) data. The spectra are normalized in velocity space, with
zero--point at $z=0.9276$. A $1\sigma$ level error is indicated by the
line at the bottom of each spectrum. The ``$\ast$'' above the
continuum indicates that the spectrum is contaminated by a known blend
at that location (see \S~\ref{sec:stisblends}). The solid lines
superimposed on the {\MgII} spectrum are the Voigt profile fit. The
vertical ticks above the fit label the locations of the individual
{\MgII} clouds obtained from the Voigt profile fit. The designated
cloud numbers are consistent with the ones listed in
Tables~\ref{tab:tab3}--\ref{tab:tab5}.}
\label{fig:data1}
\end{figure*}

\begin{figure*}
\figurenum{2} 
\plotone{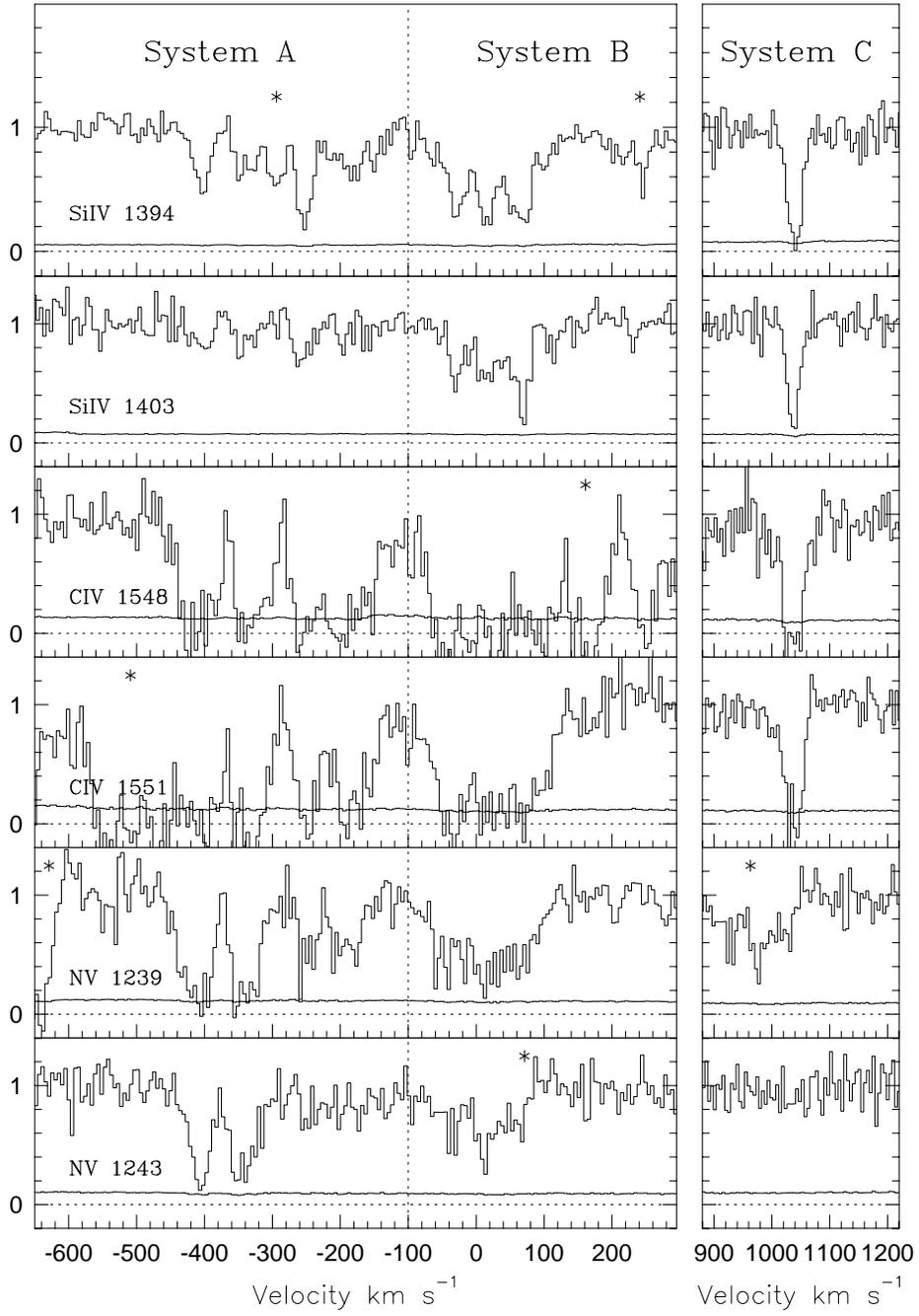}
\vglue -0.5in
\protect\caption 
{Same as Figure~\ref{fig:data1}, but for the high--ionization
transitions.}
\label{fig:data2}
\end{figure*}

The STIS  spectrum covers {\Lya}, {\SiII},  {\CII}, {\SiIII}, {\SiIV},
{\CIV}, and {\NV} transitions from  the three systems.  These data are
presented in Figures~\ref{fig:data1} and \ref{fig:data2}.  The density
of absorption  lines in the near  UV is relatively  high, resulting in
several  blends between the  three systems  and also  interlopers from
metal--line systems at other redshifts \citep[see][]{kp13}. Interloper
absorption is indicated with an ``$\ast$''.

In Table~\ref{tab:tab1}, we  present the rest--frame equivalent widths
and $3~\sigma$  limits for the  ultraviolet absorption lines  shown in
Figures~\ref{fig:data1}  and \ref{fig:data2}.   Here,  we discuss  the
blends  and   interloper  identifications   in  the  order   that  the
transitions    are    presented    in   Figures~\ref{fig:data1}    and
\ref{fig:data2}.

The {\Lya} in system B is blended with Galactic {\FeII}~$\lambda$2344
at $v\simeq +100$~{\kms}.  {\SiII}~$\lambda$1260 in system B is
blended at $v\simeq +200$~{\kms} with {\SiII}~$\lambda$1304 at
$z=0.8640$.  {\CII}~$\lambda$1334 in system A is blended at $v\simeq
-350$~{\kms} with a $z=0.6568$ {\CIV}~$\lambda$1551 line, since the
{\CIV}~$\lambda$1548 absorption at the corresponding wavelength is
also found.  The {\CII}~$\lambda$1334 in system B is affected by an
unidentified blend at $v\simeq +140$~{\kms} and in system C by an
unidentified blend at $v\simeq +950$~{\kms}.  These interloping lines
could be due to {\Lya}, since the {\Lyb} absorption at the
corresponding wavelengths is detected.  Absorption at $v \sim
-300$~{\kms} in {\SiIV}~$\lambda$1394 in system A is at least
partially due to {\CIV}~$\lambda$1548 at $z=0.7338$.  The
corresponding {\CIV}~$\lambda$1551 absorption is present at $v \sim
+230$~{\kms} in the {\SiIV}~$\lambda$1394 absorption of system B.  The
self--blending of the {\CIVdblt} is severe; {\CIV}~$\lambda$1548 from
system B is strongly blended with {\CIV}~$\lambda$1551 from system A.
At $v\simeq -600$~{\kms}, {\NV}~$\lambda$1239 from system A is
contaminated with Galactic {\FeII}~$\lambda$2383.  Also,
{\NV}~$\lambda$1243 in system B is blended with {\NV}~$\lambda$1239
from system C.

\subsubsection{Far Ultraviolet Wavelengths}
\label{sec:fosblends}

\begin{figure*}
\figurenum{3} 
\plotone{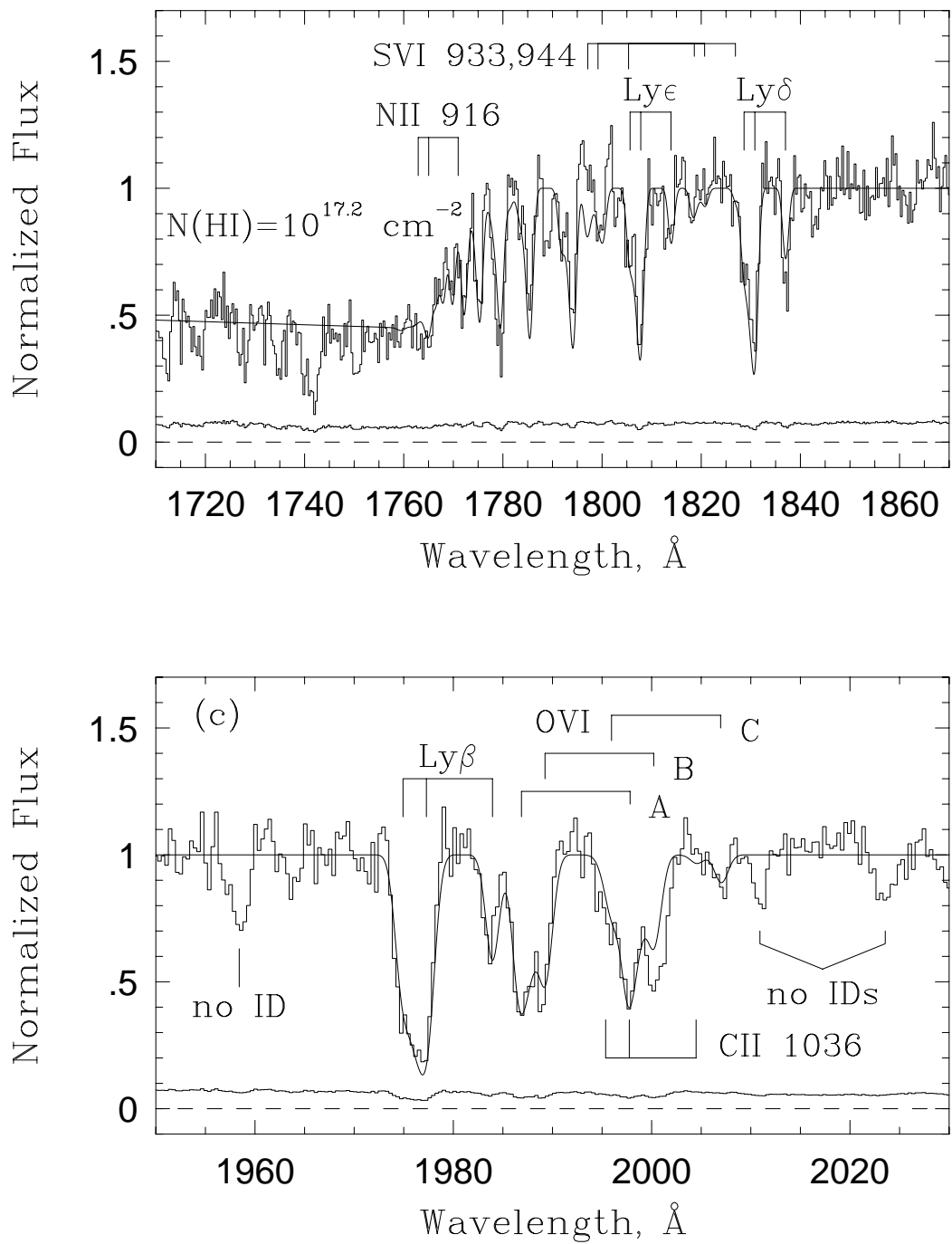}
\vglue -0.3in 
\protect\caption 
{A partial Lyman limit break, covered by the FOS spectrum, is
displayed in wavelength space above. The higher order Lyman series and
the {\OVI} profiles are also shown. An example of an acceptable model
for the three systems (the solid curve) is superimposed on the data
(histogram).  The parameters of this model are listed in
Tables~\ref{tab:tab3}--\ref{tab:tab5}.}
\label{fig:fos}
\end{figure*}

We show two regions of  the FOS spectrum in Figure~\ref{fig:fos}.  The
upper panel shows the  partial Lyman--limit break and high--order {\HI} lines.
Also covered are the {\NII}~$\lambda$916 transition and the {\SVIdblt}
doublet.   The ``three--armed''  ticks  identifying these  transitions
give the locations  of systems A, B, and C,  respectively from blue to
red.   The solid  curve  through the  data  is synthesized based  upon
our  models
(discussed  below).   Based on the partial Lyman--limit break,
\citet{q1206}  determined  that the effective optical depth,
$\tau({\HI})\simeq 1$, which arises predominately from clouds eight and
nine in system B.

The lower panel in  Figure~\ref{fig:fos} shows the {\OVIdblt}, {\Lyb},
and    {\CII}~$\lambda$1036   transitions.     Various    blends   are
apparent.   {\OVI}~$\lambda$1032   in  system   C   is  blended   with
{\CII}~$\lambda$1036 in  system A, as  well as with  some unidentified
absorption, possibly  a {\Lya} forest  line.  The {\OVI}~$\lambda$1038
line in system A is blended with {\CII}~$\lambda$1036 in system B, and
{\OVI}~$\lambda$1038  in system  B is  blended with  some unidentified
absorption,  possibly  another   {\Lya}  forest  line  \citep[see][for
further details]{q1206}.

\section{Galaxies in the QSO Field}
\label{sec:galaid}

\subsection{WIYN Image}
\label{sec:image}

In   Figure~\ref{fig:image},   we   present  a   $15{\arcsec}   \times
15{\arcsec}$ segment of the WIYN  $i$--band image centered on the QSO.
Four galaxies,  labeled G1--G4, were detected at  $9.7$, $8.6$, $9.7$,
and  $14.5$\arcsec,  respectively.   Morphologically,  G1  appears  to
resemble a  warped spiral  galaxy and G2  is consistent with  being an
early--type galaxy.  G3 and G4  are somewhat amorphous.  Since we lack
standard  flux calibrations  to perform  absolute  photometry directly
from  our  images,  we  obtained  relative  magnitudes  for  the  four
candidate  absorbing galaxies.   Relative to  G1, we  find that  G2 is
$0.3$ magnitudes  fainter, G3 is  $1.9$ magnitudes fainter, and  G4 is
$2.8$ magnitudes fainter (with a $\sim$ 0.1 magnitude uncertainty).

Assuming,  for the moment,  that G1--G4  are at  $z \simeq  0.93$, the
corresponding  impact  parameters  are   $43$,  $38$,  $43$,  and  $65
h^{-1}$~kpc ($q_0  = 0.5$).  For  reference, the impact  parameters of
galaxies  identified  as  {\MgII}  absorbers  with  $W_{r}(2796)  \geq
0.3$~{\AA} are observed to be  distributed within $\sim 40$~kpc of the
QSO line  of sight \citep{bb91,s95}.  Again,  assuming $z\simeq 0.93$,
we  obtain  rough  absolute  magnitudes  by  scaling  the  results  of
\citet{kirhakos}, who  report a Gunn--$g$ magnitude of  $21.5$ for G1.
Ignoring color  terms, we  roughly estimate luminosities  of $2L_K^*$,
$L_K^*$,  $0.2L_K^*$,  and $0.1L_K^*$  for  G1--G4, respectively.   In
Table~\ref{tab:tab2}, we  have listed these  inferred properties based
upon the assumption of $z\simeq 0.93$ for the four galaxies.

Based upon the results of \citet{sdp94} and \citet{s95}, it is
statistically favorable that one or more of these galaxies hosts
{\MgII} absorption.  Using the criteria for associating a galaxy with 
{\MgII} absorption listed by \citet{s95}, the two most luminous
galaxies (G1 and G2) effectively lie at the luminosity dependent
boundary, $38 h^{-1} (L_K/L_K^*)^{0.15}$~kpc, within which we expect
{\MgII} absorption having $W_{r}(2796) \geq 0.3$~{\AA}.  It has not
yet been directly determined whether weaker {\MgII} absorption is
found beyond this boundary for normal, bright galaxies, or if there is
a sharp cut--off (however, see \citet{bowen95} and \citet{disks}).
Under the assumption that all weak {\MgII} absorbers are also
associated with normal, bright galaxies, \citet{weak1} show that
absorption with $0.02 \leq W_{r}(2796) < 0.3$~{\AA} would arise
at impact parameters between $\sim 40$ and $70 h^{-1}$~kpc.

The  PG~$1206+459$  sightline   has  several  intervening  metal--line
absorption systems  in addition to  the $z=0.93$ complex.   Indeed, we
identify seven  additional galaxies within a radius  of 40{\arcsec} of
the QSO.  Their apparent magnitudes range from 0.8 to 2.2 fainter than
G1.  Though there is a  weak redshift dependence for determining their
physical  impact parameter,  we  estimate  that they  all  lie in  the
approximate range $80$--$200$~kpc of  the sightline.  This places them
all outside  the range where  strong {\MgII} absorption  is predicted.
According  to \citet{chen01},  these  are good  candidates for  {\CIV}
absorbers, provided they are intervening the QSO.


\subsection{Redshifts of the Candidate Galaxies}
\label{sec:galaspec}

In  the CryoCam  spectra,  the  continuum of  G1  was only  marginally
detected and  no useful spectral features were present  to facilitate
measurement of a
redshift of  the galaxy.   For G2, a  strong emission line was  measured at
$7190\pm2$~{\AA}  (fixed  to vacuum  wavelengths  with a  heliocentric
correction for comparison to  the absorption redshifts).  If this line
is   {\OII}~$\lambda$3727,   the   galaxy   is  at   a   redshift   of
$0.9289\pm0.0005$.

This redshift places G2 at a velocity of $+200\pm80$~{\kms} relative
to the velocity zero point of system B.  This galaxy--absorption
velocity difference is consistent with that seen in a small sample of
{\MgII} galaxies \citep{steidel02}.  Though there is no definitive
data to identify G2 as the host galaxy for system B, we favor this
interpretation.  If system A, or system A+B were hosted by G2, the
galaxy--absorber velocity difference would extend to $750$~{\kms},
which we find to be a less likely scenario {\it if there is a
one--to--one absorber--galaxy correspondence.}

Additional information  on the galaxy  redshifts has been  reported by
\citet{thimm95},   who  published  a   Fabry--Perot  image   tuned  to
redshifted {\OII}~$\lambda$3727 at $z=0.93$.  G2 is clearly present in
the Thimm  image.  Though not  discussed by \citet{thimm95},  a visual
inspection of the image also reveals a marginal {\OII} emission signal
from  G3\footnote{G.~Thimm could  not be  contacted so  that  we could
objectively study this image.}.

In conclusion, we find that G2 is at the redshift of the
absorbers. Considering the absorption properties of system B and the
luminous properties and impact parameter of G2, we find that a system
B association is highly consistent with the population of {\MgII}
absorbers.  Based upon the Thimm data, we find (albeit less
definitively) that G3 is also located at the redshift of the
absorbers.  However, we are unable to argue a clear one--to--one
association with one of the three absorption systems.  We have no
redshift information for the bright galaxy G1, nor for the more
amorphous G4.  However, the luminosity and impact parameter of G1 are
consistent with it hosting strong {\MgII} absorption.  In fact, we
note \citep[see][]{disks} that had G1 been the only galaxy in the
field within $\sim 10${\arcsec} of the QSO, it would have been
incorporated into the ``absorbing galaxy'' sample of \citet{sdp94} and
\citet{s95} {\it even without a confirmed redshift}\footnote{Only
70\% of the \citet{sdp94} galaxies were spectroscopically confirmed.}.


\section{Methods for Modeling}
\label{sec:methods}

We  follow the  general technique  of \citet{q1206}.   Our goal  is to
place  constraints on the  chemical and  ionization conditions  of the
absorbing gas.   Our first assumption is  that the gas  is in
ionization   equilibrium,   either   photoionization  or   collisional
ionization.  Our  second  assumption is that  the gas is  in a
minimal number of ionization phases, where a phase is a set of
clouds with metallicities and ionization parameters/densities
(more on this below).

\subsection{General Procedure}

In very general terms, we proceed by first assuming the gas giving
rise to {\MgII} absorption is in photoionization equilibrium.  System
by system, we constrain the ionization conditions and metallicities of
components (or ``clouds'') 1--12, using their velocities as a
kinematic template upon which to compare all the observed transitions
in the HIRES, STIS, and FOS spectra.  As found in \citet{q1206}, we
know that the {\MgII} clouds have fairly low ionization conditions.
That is, $n_{H} \simeq 0.01$~{\cc}, with the ionization fractions of
Mg, Fe, C, and Si being dominated by singly ionized species.

These clouds  do not give  rise to the  bulk of the {\CIV}  {\NV}, and
{\OVI}  absorption  \citep[see][]{q1206}.   Thus,  we then  model  the
residual absorption from these  species (i.e.\ the observed absorption
remaining in the spectra after accounting for the absorption predicted
by  the   low--ionization  cloud   models)  as  a  separate   zone  of
high--ionization,  lower density  gas.  We  explored single  phase and
multiphase  photoionization   models,  single  phase   and  multiphase
collisional  ionization  models, and  multiphase  photo +  collisional
ionization models to describe high--ionization gas.

Because of  the excellent resolution of the  kinematics, we discovered
that we  needed to introduce  a third, intermediate--ionization phase
over two narrow  velocity intervals to account for  some of the {\CIV}
and {\SiIV}  absorption.  The  absorption in these  velocity intervals
could  not be  satisfactorily described  within the  allowed parameter
spaces of models limited to low-- and high-- ionization phases.

\subsection{Photoionization Modeling}

We used the photoionization code Cloudy, version 90.4 \citep{cloudy}.
Each model cloud has constant density and plane--parallel geometry.
An important parameter describing each cloud model is the ionization
parameter, $U$, which is defined as the ratio of the incident ionizing
photon density, $n_{\gamma}$, to the number density of hydrogen,
$n_H$.  For the photoionizing flux, we used the extragalactic
ultraviolet spectrum of \citet{haardtmadau96}.  Normalized at $z=1$ ,
the number density of ionizing photons (capable of ionizing hydrogen)
incident on the model clouds is $\log n_{\gamma} = -5.2$~[{\cc}].  For
the $z=1$ normalization of the \citet{haardtmadau96} spectrum, the
number density of hydrogen is $$\log n_H = -5.2 - \log U~{\cc}.$$  The
effects of alternative spectral shapes will be discussed in
\S~\ref{sec:specshape}. The abundance pattern was assumed to be solar
(using the standard Cloudy abundances).  Alternatives to a solar
pattern were explored as suggested by the data.

Input  parameters to  Cloudy included:  i) the  observed  or estimated
column density of a  well constrained transition, ii) the metallicity,
$Z$  (expressed in  units of  the  solar value),  iii) the  ionization
parameter, $U$, and iv) the  abundance pattern.  For each model cloud,
we record  the column  density of transitions  typically found  in QSO
absorption  line  systems that  were  covered  by  our spectra  (i.e.\
including  transitions  not  formally  detected and  not  presented  in
Figures~\ref{fig:data1}  and  \ref{fig:data2}).   We also  record  the
kinetic temperature,  $T$, which we equate with  the thermal component
of the Doppler parameter, $b_{therm}^2  = 2kT/m$, where $m$ is the ion
mass.

We  assume a  turbulent component  to the  Doppler parameter  in cases
where the model cloud temperature implies a smaller value than what is
measured in the spectra.  This turbulent component, which we assume is
identical  for all  species, was  calculated using  $b_{\rm  turb}^2 =
b_{\rm ion}^2 - b_{\rm therm,\rm ion}^2$, where ``turb'' and ``therm''
denote the  turbulent and  thermal components, respectively.   We then
have the  ingredients to estimate the ``total''  Doppler parameter for
any  observed  transitions  by   inverting  this  relation  using  the
appropriate ion mass.

We converge on the range of metallicity and ionization conditions by
an iterative process of comparing synthetic spectra generated for each
transition with column densities and temperatures of the cloud models.
These spectra were convolved with the appropriate instrumental spread
functions (using FFTs) in order to directly compare with the observed
spectra.  For the HIRES and FOS spectra, we use a Gaussian with FWHM$
= 6.6$~{\kms} and $230$~{\kms}, respectively.  For the STIS spectra,
we used the tabulated instrumental spread function appropriate for the
E230M grating as published on STIS performance web pages at {\it
www.stsci.edu}.

For  each iteration, the  synthetic spectra  are superimposed  on the
observed  spectra.  A  pixel by  pixel $\chi^2$  was calculated  as an
indicator of the quality of  the overall match.  However, the value of
$\chi^2$ was often dominated by  large values in a very small fraction
of ``bad''  pixels (i.e.\ those  having correlated noise spikes  or an
interloping  line  blend).   This  rendered any  objective  $\chi^{2}$
minimization scheme  fairly ineffective, since it yielded  a very flat
``$\chi^{2}$  landscape''.  Visual  inspection, guided  by  $\chi^{2}$
minimization, was  essential for comparing the  full kinematic profile
shapes in the synthetic spectra to the observational spectra.

As described in Appendix~A  of \citet{q1206}, the ionization parameter
was generally constrained by  the relative strength of the metal--line
transitions and  the metallicity by {\Lya}, higher  order Lyman series
lines, and Lyman break from  the FOS spectrum.  Generally, this method
yields both $\log U$ and  $\log Z$ constrained within $\simeq 0.1$~dex
(for the  assumed abundance pattern and  kinematic parameterization of
the profiles).

\subsection{Collisional Ionization Modeling}

We   used   the   collisional   ionization   equilibrium   models   of
\citet{cooling  function}.   These   models  have  a  solar  abundance
pattern.   The two  free parameters  in these  models are  the kinetic
temperature, $T$,  and metallicity, $Z$.   The ion fractions  for each
transition are a unique function of temperature from the models.

Given the metallicity, $Z$, and a measured or estimated column density
of a given  species, $N^i$, the total H  column density was determined
using $N (H) = N^i / (A*Z*r^i)$, where $A$ is the fractional abundance
of the element  relative to hydrogen in a  solar abundance pattern and
$r^i$ is the  ion fraction of the species.   Then, column densities of
various species could be calculated using $N_X^i = N (H)*A_X*Z*r_X^i$,
where $N_X^i$ is the column density of an arbitrary species.

The Doppler parameter of each ion could be obtained in the same way as
for  the  photoionization  model,   with  $T$  being  the  collisional
temperature.  In the  case where  the  observed Doppler  parameter of  the
optimized  species was  consistent with  $b^2 =  2kT/m$,  then $b_{\rm
turb}=0$~{\kms} was used  for all ions. Similar to  the procedure for
photoionization  modeling, we  generated synthetic  model  spectra and
superimposed them upon the data for comparison.

\subsection{Defining and Modeling Phases}

A  main goal of  our modeling  is to  parameterize the  gas in  as few
ionization  ``phases'' as  possible.   In practice,  the data  require
at least three phases.

\subsubsection{Low Ionization}
\label{sec:low}

We define the term ``{\MgII} phase'' to denote the ionization
conditions which predominantly (or fully) describe the density and
ionization balance of the gas giving rise to {\MgII} absorption.  For
our models, this phase is characterized by number densities in the
relatively narrow range $-2.7 \la \log n_{H} \la -2.0$~{\cc},
ionization parameters in the range $-3.2 \la \log U \la -2.5$ and
hydrogen Doppler parameters of $15 \la b({\rm H}) \la 25$~{\kms}.

The metallicity and ionization parameter were explored over the range
$\log Z$ from $-2$ to $0.5$ in intervals of $0.1$~dex and $\log U$
from $-3.5$ to $0$ in intervals of $0.1$~dex.  To begin, we assumed
that for each system, all the {\MgII} clouds in that system have the
same metallicity and abundance pattern.  This assumption is likely to
not reflect reality (e.g. \citet{ganguly}), but in the absence of
resolved higher order {\HI} transitions, there are virtually no
constraints on the cloud to cloud metallicities.  However, we explored
variations in the abundance pattern as suggested by the data.

We ``optimized''  each cloud model  on the {\MgII} column  density for
trial $U$ and $Z$.  Recall that column densities of the {\MgII} clouds
were obtained from the  Voigt profile fits \citep{cv01,q1206}.  In its
optimized mode,  Cloudy adjusts $N({\HI})$, until  the $N({\MgII})$ of
the  model  cloud differs  from  the measured  value  by  less than  a
specified percent difference (we chose $5$\%).

\subsubsection{High Ionization}
\label{sec:high}

Analogous to ``{\MgII}  phase'', we use the term  ``diffuse phase'' to
denote the gas with  significantly lower density and higher ionization
balance.  The  conditions describing this phase  were less constrained
than  those describing  the  {\MgII} phase.   For our  photoionization
models, the diffuse phase is  characterized by number densities in the
range $-4.7 \la \log  n_{H} \la -3.6$~{\cc}, ionization parameters in
the range $-1.6 \la \log U \la -0.5$ and hydrogen Doppler parameters
in the range $20 \la b({\rm H}) \la 50$~{\kms}.

It was difficult to obtain a velocity template for the diffuse phase.
The {\NV} doublet in system A had sufficient velocity structure (from
$-450$ to $-150$~{\kms}) and signal--to--noise ratio to perform a
formal automated VP fit.  However, for system B, we had to use a
``by--hand'' method PROFIT to estimate VP fit parameters.  The {\NV} and
{\CIV} profiles provided the best constraints for ionization
parameters of clouds in this phase.  We followed the procedures
described in \S~\ref{sec:low} to obtain a synthetic spectrum for each
of the other observed transitions.  The ionization parameter was tuned
to attempt to produce {\CIV}, {\NV}, and {\OVI} in this phase, without
overproducing {\SiIV} and other intermediate--ionization transitions.
The metallicity was constrained mainly by the {\Lya} profile.

\subsubsection{Intermediate Ionization}
\label{sec:inter}

Procedurally, we approach each system by first constraining the
conditions of the {\MgII} phase, and then the diffuse phase.  However,
unaccounted absorption still appeared in some individual transitions
at certain velocities, particularly in transitions with
intermediate--ionization state.  Examples are the {\CIV} profiles at
$v \sim -157$~{\kms} and the {\SiIV} profiles at $v \sim
+66$~{\kms}. Therefore, additional clouds with
intermediate--ionization states were considered.

Both photoionization and collisional  ionization were explored. In the
case  of photoionization, Cloudy  was run,  optimized on  the measured
{\CIV} or {\SiIV} column  density.  The range of ionization parameters
(for  photoionization) and  temperatures (for  collisional ionization)
were    found    so   that    no    other    transitions   would    be
overproduced.  Metallicity was  constrained by  {\Lya} and  the higher
order  Lyman  series  lines.    Additional  details  of  the  modeling
techniques were identical to  the descriptions in \S~\ref{sec:low} and
\S~\ref{sec:high}.


\section{Constraining the Absorber Properties}
\label{sec:results}

Specific constraints on the phase structure of systems A, B, and C are
presented here.  In the text, we discuss what ranges of parameters are
consistent with the data, considering scenarios of photoionization and
of collisional ionization for the various required phases.
For each photoionized phase we discuss
constraints on the ionization parameter and metallicity, and mention
the transitions that allow us to place these constraints. For
collisionally ionized phases, we similarly place constraints on the
temperature and metallicity.

In Tables~\ref{tab:tab3}--\ref{tab:tab5}, we give more detailed
information for examples of
acceptable photoionized and collisionally ionized model phases.
These can be considered loosely as ``best models'', intermediate in the
allowed ranges.  For each model, we list metallicities,
ionization parameters, densities, sizes, and temperatures of model
clouds for systems A, B, and C, respectively.  We also list the column
densities, and Doppler parameters of selected species.

In Figures~\ref{fig:fitA}--\ref{fig:fitC} we present the data with the
synthetic spectra from the ``best models'' superimposed.  The clouds
are identified by number above the continuum of the
{\MgII}~$\lambda$2796 profile.  The synthetic spectra through the
HIRES data ({\FeII} and {\MgII}) are derived from the Voigt profile
fits.  For the STIS spectra, the solid curves through the data are the
full model, with contributions from the {\MgII}, intermediate-- and
high--ionization phases.  The dotted--line synthetic spectra are the
contribution to the absorption profiles from the {\MgII} phase only.
The dashed--dot line synthetic spectra are for the
intermediate--ionization phase, and the dashed-line spectra are for
the high--ionization phase.

\subsection{Constraining System~A Properties}
\label{sec:A}

\subsubsection{\MgII~Phase}
\label{sec:AMgII}

The ionization parameters, $\log U$, of the six {\MgII} clouds of
system A were constrained by optimizing on the {\MgII} column density
from the VP fit, and adjusting $\log U$ to best fit other low-- and
intermediate--ionization transitions.  In order to minimize the number
of phases, we assume that as much {\SiIV} as possible is produced in
the {\MgII} clouds.  (The {\SiIV} components are too narrow to be
produced in the high--ionization phase.)  Thus, ionization parameters
of the {\MgII} clouds are mainly constrained by the fit to
{\SiIV}~$\lambda$1403\footnote{As mentioned in
\S~\ref{sec:stisblends}, {\SiIV}~$\lambda$1394 is contaminated with
{\CIV}~$\lambda$1548 from a system at $z=0.7738$}. The ionization
parameters for the six clouds, {\MgII}$- \rm 1$ through {\MgII}$- \rm
6$, are within the range, $-2.8 \la \log U \la -2.5$, as listed in
Table~\ref{tab:tab3}.

These values of $\log U$ produce absorption consistent with the data
in other low--ionization transitions, such as {\SiII} and {\SiIII},
and with the limit set by the undetected {\FeII}.  (see the dotted
curve in Figure~\ref{fig:fitA}).  However, {\CII}~$\lambda$1334
appears to be underproduced by such models.  This is partly due to
contamination ($-360 \la v \la -340$~{\kms}) by a blend with
{\CIV}~$\lambda$1551 at $z=0.6568$ (see \S~\ref{sec:stisblends}).
However, the underproduction at $-420 \la v \la -400$~{\kms} still
remains unexplained, and suggests an abundance pattern that deviates
from solar.  In order to fit the {\CII} spectra we would need to
increase carbon abundance by $0.4$--$0.8$~dex in the various clouds.

Using $-2.8 \la \log U \la -2.5$ and a solar abundance pattern, with
the assumption that the six {\MgII} clouds have the same metallicity,
a super--solar metallicity of $\log Z \simeq 0.5$ produced the best
fit to the Lyman series lines ({\Lyb} and the higher order lines; see
Figure~\ref{fig:fos}).  To achieve a comparable fit with $\log Z =
0.0$, we would need to increase the abundance of magnesium by 1.0~dex
(the usual ``trade--off'' between metallicity and abundance pattern
does not apply for supersolar metallicities).  If the increased
abundance of magnesium was due to an $\alpha$--group enhancement, then
silicon would be increased as well, but carbon would not
\citep{jtl96}.  With carbon decreased relative to magnesium by $\sim
1.0$~dex, the discrepancy between $\log Z = 0.0$ model and the {\CII}
profiles would be considerably more extreme.

We therefore favor models with $\log Z \simeq 0.5$ and roughly solar
abundance patterns.  For this $\log Z$, the blue side of the {\Lya}
profile is underproduced.  A $\log Z \simeq 0.0$ would produce an
adequate fit to {\Lya}, but would severely overproduce the other
Lyman series lines.  A high--ionization phase capable of fitting the
blue side of the {\Lya} profile will be discussed in \S~\ref{sec:Adiff}.

The column densities of {\HI} for these {\MgII} clouds fall within the
range $14.4 \la \log N ({\HI}) \la 15.1$~[{\cmsq}].  Thus, this
low--ionization phase in system A makes an insignificant contribution
to the observed partial Lyman limit break, which requires an {\HI}
column density of $\log N ({\HI}) \sim 17$~[{\cmsq}].  Given the
number density $n_H$, 
$$\log n_H = -5.2 - \log U,$$ 
and the total column density of hydrogen $N_{\rm tot}({\rm H})$, for
$\log Z = 0.5$ the sizes of the six {\MgII} clouds are within
$5$--$20$~pc, using $l$ = $N_{\rm tot}({\rm H})$ / $n_H$.

\subsubsection{Diffuse~Phase}
\label{sec:Adiff}
The strong absorption in the {\CIV}, {\NV}, and {\OVI} profiles (see
Figure~\ref{fig:fitA}), which is not fully accounted for by the
{\MgII} phase, requires a diffuse phase.  Also, the blue side of the
{\Lya} profile could not be fit by the {\MgII} cloud contribution
without overproducing {\Lyb} and the higher order Lyman series lines.

We explored both the possibility of photoionization and of collisional
ionization for the diffuse phase.  For the photoionization models, we
optimized on the seven {\NV} column densities given by AUTOVP MINFIT.
The Doppler parameters from the VP fit were in the range $3 \la b({\rm
N}) \la 21$~{\kms}, as listed in Table~\ref{tab:tab3} under ``Case A:
Photoionization''.  Collisional ionization of {\NV} would require
$b({\rm N}) \ge 15$~{\kms}.  Therefore, for collisional ionization
models we adjusted $b({\rm N})$ and $N({\NV})$ of the VP fit
components using PROFIT so that all of the Doppler parameters exceeded
$15$~{\kms}, as listed in Table~\ref{tab:tab3} under ``Case B:
Collisional Ionization''.  This fit for a collisional ionization phase
case was plausible, but not unique.

For the case of photoionization (Case A in Table~\ref{tab:tab3}),
{\CIV}, {\NV}, and {\OVI} could arise in the same phase.  Optimizing
on $N({\NV})$, we found that the ionization parameters of the
high--ionization phase clouds are constrained to be within the range
$-1.5 \la \log U \la -0.6$.  Clouds {\NV}$- \rm 1$ and {\NV}$- \rm 3$
have higher ionization parameters ($\log U \simeq -0.6$) and are the
primary producers of {\OVI}. The other five diffuse clouds, with lower
ionization parameters ($-1.5 \la \log U \la -1.2$), give rise to the
{\CIV} absorption.

To derive constraints on metallicity for the photoionized case, we
assume that the seven clouds have the same metallicity and that they
have a solar abundance pattern. A metallicity of $\log Z
\simeq 0.5$ is required to fit the blue side of the {\Lya} line.  This
{\Lya} absorption is produced dominantly by cloud {\NV}$- \rm 2$ (see
Table~\ref{tab:tab3}), so it is important to note that the
metallicities of the individual diffuse phase components are not
well--constrained.  The inferred metallicity of $\log Z \simeq 0.5$
could be reduced to solar if nitrogen was enhanced by $\sim
0.5~$dex. The properties of the seven diffuse, photoionized clouds
(with sizes $1$--$65$~kpc) are listed in Table~\ref{tab:tab3}.

The case of collisional ionization is more complicated.  The {\CIV},
{\NV}, and {\OVI} cannot be produced in a single temperature
collisionally ionized model.  We explored the scenario (Case B in
Table~\ref{tab:tab3}) in which {\CIV} is produced in the same
collisionally ionized clouds as {\NV}.  The {\CIV} and {\NV} profiles
require the temperatures of the seven clouds to be within the range
$5.14 \la \log T \la 5.18$~[K].  Since {\OVI} is not produced within
this temperature range, an additional phase is needed to give rise to
the observed {\OVI} absorption. The physical parameters of this
additional phase are not well constrained, due to the lack of any
resolvable feature in the low--resolution spectrum of
{\OVIdblt}. Therefore, we assume a single cloud, centered on the
{\OVI} absorption at the redshift $z \sim 0.92545$, with $\log N
({\OVI}) \sim 15.0$~[{\cmsq}] and $b ({\rm O}) \sim 80$~{\kms}. This
additional {\OVI} cloud could be either photoionized or collisionally
ionized. For photoionization (cloud {\OVI}$- \rm ph$), an ionization
parameter of $\log U \sim -0.6$ produces the observed {\OVI}. This
ionization parameter is high enough not to give rise to any
lower--ionization transitions and low enough that the cloud remains
Jeans stable.  For collisional ionization (cloud {\OVI}$- \rm co$), a
temperature of $\log T \sim 5.5$~[K] would not overproduce any other
transitions. Assuming that the metallicity of the {\OVI} cloud to be
the same as the clouds that in which {\NV} arises, a metallicity of
$\log Z \simeq 0.4$ is required to fit the blue side of the {\Lya}
profile.

In principle, a model in which {\NV} and {\OVI} are produced in a
higher temperature collisionally ionized phase, and the {\CIV} in a
lower temperature collisionally ionized phase is also possible.
However, it is difficult to place detailed constraints on this phase
because of the blended components in the saturated {\CIV} profiles.

We conclude that {\NV} could be produced in a collisionally ionized
phase, but that in this case an extra phase would need to be
introduced to explain either the {\CIV} or the {\OVI} absorption.
Since a single diffuse photoionized phase can be made consistent with
all three high--ionization transitions, we favor a photoionized model.
The dashed lines in Figure~\ref{fig:fitA} represent the contribution of
the photoionized diffuse phase.  In \S~\ref{sec:coso6} we consider
the differences between the {\OVI} profiles for photoionization and
collisional ionization models.

\subsubsection{Intermediate~Phase}
\label{sec:Aintermediate}

Neither the {\MgII} phase or any of the diffuse phase models could
fully account for the absorption to the red of the {\CIV} profiles (at
$v \sim -160$~{\kms}), as shown in Figure~\ref{fig:fitA}.  An extra
component, either photoionized or collisionally ionized, thus seems
necessary. For the case of photoionization, the ionization parameter
is constrained to be $\log U \simeq -2.0$. For this ionization
parameter, metallicity of $\log Z \simeq -1.0$ is required to fill in
the residual in the {\Lya} profile, at $v\sim-125$~{\kms}.  Also, a
smaller value of $\log Z$ would overproduce the higher order Lyman
series lines. The contributions of this photoionized component to
{\CIV} and {\Lya} are shown as dashed--dotted curves in
Figure~\ref{fig:fitA}. The size of the cloud is $\sim 5$~kpc.
Alternatively, for the case of collisional ionization, the temperature
is $\log T \sim 5.0$~[K], constrained by {\SiIV}, {\CIV}, and {\NV}.
In this case, the metallicity of is constrained to be $\log Z \la
-2.0$ to fit the residual in the {\Lya} profile.

\subsection{Constraining System~B Properties}
\label {sec:B}
\subsubsection{\MgII~Phase}
\label {sec:BMgII}
The strong, blended {\MgII} profile in system B was fit with five
narrow components ($b \sim 5$--$17$~{\kms}), spread over $\sim 200~$
{\kms} in velocity space (see Figure~\ref{fig:fitB}).  In
Table~\ref{tab:tab4}, the column densities and Doppler parameters from
the VP fits to {\MgIIdblt} are listed for these clouds, designated as
{\MgII}$- \rm 7$ through {\MgII}$- \rm 11$.  Clouds {\MgII}$- \rm 8$,
{\MgII}$- \rm 9$, and {\MgII}$- \rm 11$ have detected {\FeII}
absorption. The ratio $N ({\FeII})$/$N ({\MgII})$ for those
three clouds thus constrains their ionization parameters to be $\log U
\simeq -3.2$, $-3.0$, and $-3.2$ \citep{q1206}.  These ionization
parameters for clouds {\MgII}$- \rm 8$ and {\MgII}$- \rm 9$ are
consistent with the absorption seen in {\SiII}, {\SiIII}, and
{\SiIV}. However, cloud {\MgII}$- \rm 11$ does not produce a
sufficient amount of {\SiIV} to match the observed profile. An
additional phase is needed, as described in\S~\ref{sec:Bintermediate}.
For the other two clouds, {\MgII}$- \rm 7$ and {\MgII}$- \rm 10$, the
ionization conditions are determined by the strength of {\SiIV} to be
$\log U \simeq -2.5$ and $-2.9$, assuming that {\SiIV} is produced in
the low--ionization phase. These values produce models consistent with
the {\SiII} and {\SiIII} profiles, and with the {\FeII} limit.
Figure~\ref{fig:fitB} shows the contribution from clouds {\MgII}$- \rm
7$ through {\MgII}$- \rm 11$ as a dotted line superimposed on various
observed profiles.

The five {\MgII} clouds are assumed to have the same metallicity, but
the metallicity is primarily constrained by the cloud(s) that produces
the strongest {\HI} absorption.  The strongest constraint on this
metallicity comes from the partial Lyman limit break that is detected
in the FOS spectrum.  A small uncertainty ($\pm0.1$~dex) arises
because of the subjectivity in the continuum fit to the FOS spectrum.
Under the assumption of constant metallicity for the five {\MgII}
clouds, $\log Z \simeq 0.0$ produces the best model. Clouds {\MgII}$-
\rm 8$ and {\MgII}$- \rm 9$ have {\HI} column densities of $\log
N({\HI}) \simeq 16.7$~[{\cmsq}] and $\log N({\HI}) \simeq
16.5$~[{\cmsq}] respectively, and are the major contributors to the
break.  However, for $\log Z \simeq 0.0$, the clouds in this {\MgII}
phase do not produce sufficient absorption to account for the entire
equivalent width of the {\Lya} absorption feature\footnote{As
mentioned in \S~\ref{sec:stisblends}, {\Lya} is contaminated by
Galactic {\FeII}~$\lambda$2344.  Based on the simultaneous VP fit
performed on the other uncontaminated Galactic {\FeII} profiles, a
synthetic Galactic {\FeII}~$\lambda$2344 spectrum was produced and
included in the model of {\Lya}.}. The contribution of the {\MgII}
clouds to the {\Lya} absorption can be seen as the dotted curve on the
{\Lya} profile in Figure~\ref{fig:fitB}. For the derived metallicity
and ionization parameter constraints, the five {\MgII} clouds have
sizes of $30$--$200$~pc.

\subsubsection{Diffuse~Phase}
\label {sec:Bdiff}
In Figure~\ref{fig:fitB}, it is shown that the {\MgII} phase
makes only an insignificant contribution to {\CIV} and {\NV}.  The
{\OVI} is covered only in the low--resolution FOS spectrum, but it is
quite strong ($W_r({\OVI} 1032) \simeq 0.5~{\AA}$) for this system so
it also could not be produced by the {\MgII} phase. Also, there is no
structure apparent in the broad {\CIV} and {\NV} profiles. These facts
suggest a highly ionized, diffuse phase.

A broad VP component is placed at the velocity of cloud {\MgII}$- \rm
10$ so that it is centered on the smooth {\NV} absorption feature.
Although the data are noisy, the best single--cloud VP fit to
{\NVdblt} was obtained using PROFIT, with $\log N({\NV}) \simeq
14.3$~[{\cmsq}] and $b ({\rm N}) \simeq 50$~{\kms} (cloud {\NV}$- \rm
9$).  The {\CIV} profile is saturated so a VP fit could not be
performed, but an additional narrow component, {\CIV}$- \rm 8$, with
$b ({\rm C}) \sim 14~${\kms}, was added in order to account for the
absorption seen at $v \sim -40$~{\kms} in the {\CIV}~$\lambda$1551
profile (Figure~\ref{fig:data2}). We considered whether photoionized
and/or collisionally ionized gas at these two velocities ($v \sim -40$
and $+31$ {\kms}) could account for the {\CIV}, {\NV}, and {\OVI}
absorption.

For a photoionized model, an ionization parameter of $\log U \simeq
-1.6$ for both components produces the amount of absorption observed
in the high--ionization transitions.  The exception to this is
{\NV}~$\lambda$1243, which is blended with {\NV}~$\lambda$1239 from
system C.  The fit to {\NV}~$\lambda$1239 for system B, in
Figure~\ref{fig:fitB}, is adequate.  However, as shown in
Figure~\ref{fig:cos}, the combined model for systems B and C produces
excessive absorption at $\sim +1000$~{\kms} in the {\NV} profile.  If,
by changing model parameters, the absorption is reduced to fit the
system B {\NV}~$\lambda$1243 profile, {\NV}~$\lambda$1239 and
{\CIVdblt} would be underproduced. It is unclear how to resolve this
discrepancy if it is still present in higher S/N data.

With the assumption that both of the broad component clouds have the
same metallicity, a value of $\log Z \simeq -0.6$ is required to fit
{\Lya}, which could not be adequately produced in the {\MgII} clouds.
The higher order Lyman series lines in the FOS spectrum are consistent
with this metallicity value for the diffuse phase clouds.  Therefore,
the diffuse phase is constrained to have a somewhat lower metallicity
than the {\MgII} phase. The two diffuse components, {\NV}$- \rm 9$ and
{\CIV}$- \rm 8$, do not contribute significantly to the Lyman limit
break, due to their lower column densities ($\log N({\HI}) \simeq
15.8$~[{\cmsq}] and $\log N({\HI}) \simeq 15.0$~[{\cmsq}]). The
derived sizes of the two clouds are $\sim 54$ and $\sim 8$~kpc,
respectively.

A collisionally ionized component with a single temperature could not
simultaneously produce the observed {\CIV}, {\NV}, and {\OVI}
absorption. To minimize the overall number of clouds, we favored the
model with a single photoionized, diffuse phase. However, we explored
the possibility of a contribution from an additional collisional
phase. If {\NV} and {\OVI} arise in a collisional component, {\CIV}
would have to be produced in a photoionized phase with such a low
ionization parameter that it does not give rise to {\NV} and
{\OVI}. However, this would overproduce {\CIII}.  In order not to
overproduce {\CIII}, the photoionized phase has to be so highly
ionized that {\OVI} is fully produced.  Therefore, even an additional
collisionally ionized phase is ruled out in this system.

\subsubsection{Intermediate~Phase}
\label {sec:Bintermediate}
With an ionization parameter of $\log U \simeq -3.2$, which is
constrained by the detected {\FeII} absorption, cloud {\MgII}$- \rm 11$
does not produce the observed {\SiIV} at $v=66$~{\kms}.  Also, the diffuse
component that gives rise to the majority of {\CIV}, {\NV}, and {\OVI}
is much broader than the {\SiIV} profile. An additional component,
with $b ({\rm Si}) \sim 10$~{\kms}, is superimposed on cloud 11 to
account for the observed {\SiIV} profile. This component could either
be photoionized (cloud {\SiIV}$- \rm ph$)
or collisionally ionized (cloud {\SiIV}$- \rm co$). For the case of
photoionization, the ionization parameter is determined to be $\log U
\simeq -2.5$, constrained by {\CIV} and {\SiIII}. This component would
have a size of $\sim 2$~kpc. For the case of collisional ionization,
the temperature is determined to be $\log T\sim 4.8$~[K] (near peak
production for {\SiIV}), in order not to overproduce any other
transitions.

\subsection{Constraining System~C Properties}
\label {sec:C}
\subsubsection{\MgII~Phase}
\label {sec:CMgII}
A single, unresolved {\MgII} cloud (cloud {\MgII}$- \rm 12$) with $b
({\rm Mg}) \sim 7.5$~{\kms} is separated by $\sim 1000$~{\kms} from
system B, as seen in Figure~\ref{fig:fitC}.  {\SiIV} and the strongest
component of absorption in {\CIV} are aligned in velocity space with
{\MgII}.  However, the {\Lya} profile is not centered on the {\MgII}
cloud, and the {\CIV} profile shows evidence for an additional
``offset'' cloud at $v=+1005$~{\kms} (cloud {\CIV}$- \rm off$).
There is strong {\OVI} absorption from system C detected in the FOS
spectrum (see Figure~\ref{fig:fos}), however {\NV}~$\lambda$1243 is
only detected at the $2.7\sigma$ level in the STIS spectrum.
(The {\NV}~$\lambda$1239 is blended with {\NV}~$\lambda$1243 from
system B as discussed in \S~\ref{sec:stisblends} and \S~\ref{sec:Bdiff}.)

We favor a two--phase model to produce the absorption centered on the
{\MgII} cloud, one which produces absorption in transitions from
{\MgII} to {\CIV}, and another which gives rise to the {\OVI}
absorption (discussed in \S~\ref{sec:Cdiff}).  However, we first
consider the possibility that one phase of gas can give rise to
absorption in all these transitions.  With an ionization parameter
$-1.8 \la \log U \la -1.5$, both {\SiIV} and {\CIV} can be produced in
the same phase as {\MgII}, as can the {\SiIII}~$\lambda$1207.  The
{\OVI}~$\lambda$1038 ({\OVI}~$\lambda$1032 is blended; see
\S~\ref{sec:fosblends}) can also be produced in this {\MgII} cloud.
However, a much lower ionization parameter ($\log U \sim -3.0$) is
required to give rise to the observed {\SiII}~$\lambda$1260
absorption.  Furthermore, {\NV}~$\lambda$1243 would be severely
overproduced by this model. To achieve consistency, the abundance of
nitrogen would need to be decreased by $\sim 2$~dex relative to other
elements. Although nitrogen is deficient by $\sim 1$~dex in dwarf
galaxies \citep{mateo}, the required $2$~dex adjustment is clearly
extreme.  The one--phase model would have a super--solar metallicity
of $0.2 \la \log Z \la 0.4$ in order to fit the red wing of {\Lya}.
The size of this single {\MgII} cloud would be $10$--$160$~kpc.
However, we conclude that the problems with {\SiII} underproduction
and {\NV} overproduction are too severe, and that a one--phase model
is excluded.

In the two--phase scenario, the {\MgII} phase has a lower ionization
parameter so that {\NV} is not severely overproduced, but a separate
high--ionization phase is needed to account for the {\OVI}
absorption. The ionization parameter of the {\MgII} phase is
constrained to be $\log U \simeq -2.0$. For a lower $U$, {\CIV} would
be underproduced. However, at this ionization parameter, {\SiIV} is
not sufficiently produced. This, together with the underproduction of
{\SiII}~$\lambda$1260, suggests the need for an enhancement of
silicon.  A $0.4$~dex silicon enhancement brings the $\log U \simeq
-2.0$ model into agreement with the {\SiII}~$\lambda$1260 and
{\SiIVdblt} profiles. In this case, the less severe overproduction of
{\NV} can be resolved by a $0.4$~dex decrease of nitrogen. At $\log U
\simeq -2.0$, a metallicity of $\log Z \simeq 0.1$ matches the red
wing of the {\Lya} profile.  This model is indicated by the dotted
curve on Figure~\ref{fig:fitC}.  The size of the {\MgII} cloud in this
two--phase scenario is $\sim 4$~kpc.

Of the two approaches that we have explored for the {\MgII} phase, we
favor the two--phase scenario. The one--phase scenario is consistent
with our philosophy of minimizing the number of phases. However, a $2$
dex deficit of nitrogen is unlikely and it seems contrived to
assume that the {\SiII}~$\lambda$1260 profile is contaminated by a blend. The
two--phase scenario requires only a modest abundance pattern
adjustment (a $0.4$~dex deficit for nitrogen and a $0.4$~dex
enhancement for silicon).  This adjustment is reflected in the
model results summarized in Table~\ref{tab:tab5}.

\subsubsection{Diffuse~Phase}
\label {sec:Cdiff}
A high--ionization phase is required in the two--phase scenario in
order to fit the {\OVI}.  However, the physical parameters of this
phase are poorly constrained due to the unresolved {\OVI}
profiles, covered in the low--resolution FOS spectrum. Both
photoionization and collisional ionization could contribute to this
phase.

For the case of photoionization, a grid with $10 \la b({\rm O}) \la
80$~{\kms} was constructed, with various $\log Z$ and $\log U$
values. The ionization parameter is roughly constrained to be $-1.0
\la \log U \la -0.5$.  If the ionization parameter is lower than this,
{\NV} will be overproduced; if it is any higher, the cloud will not be
Jeans stable.  For $b({\rm O})=80$~{\kms}, {\NV} would be overproduced
unless the ionization parameter is $\log U \ga -0.5$. The metallicity
would also have to be $\log Z \ga -0.5$, in order not to overproduce
{\Lya}. As $b$ decreases, both the metallicity and the ionization
parameter can be lower. For example, for $b({\rm O})=10$~{\kms}, the
metallicity could reach $\log Z \simeq -1.5$ if $\log U \ga -1$.

For the case of collisional ionization, a similar grid was
constructed, with $10 \la b({\rm O}) \la 80$~{\kms} and with various
$\log Z$ and $\log T$ values.  Compared to photoionization, the
profiles are less sensitive to change in $b$ parameter and
metallicity.  However, the temperature is well constrained. A
stringent lower limit of $\log T \ga 5.4$~[K] prevents the
overproduction of {\NV} and {\Lya}, and a rough upper limit of $\log T
\la 6$~[K] prevents the overproduction of {\Lya}. For each
temperature, the assumption of pure thermal broadening provides as a
lower limit on the $b$ parameter. For example, at $\log T = 5.4$~[K],
$b({\rm O}) \ga 16$~{\kms} and at $\log T = 6$~[K], $b({\rm O}) \ga
32$~{\kms}.

Regardless of the ionization mechanism of the diffuse phase, {\SiIV}
is not significantly produced. Therefore, the assumption that {\SiIV}
arises in the {\MgII} clouds is valid for any two--phase solution. A
three--phase scenario, with the lowest--ionization phase giving rise
to the majority of {\MgII}, the intermediate--ionization phase to the
majority of {\SiIV}, and the highest--ionization phase to {\OVI}, is
excluded. The intermediate phase would either have a high enough
ionization parameter so that {\NV} would be severely overproduced, or
it would have a low enough ionization parameter so that it gives rise
to the majority of the {\MgII}.  In the latter case, the original
{\MgII} phase is redundant.

\subsubsection{Offset~Phase}
\label {sec:Coffset}
The {\Lya} profile is centered at a different velocity than the
{\MgII} ($ \sim 37~${\kms} to the blue). Also, in order to produce the
``wing'' at the blue edge of the {\CIV} profiles, a VP fit required an
offset cloud with $b ({\rm C}) \sim 15.5$~{\kms} at $v = +1005$~{\kms}
(cloud {\CIV}$- \rm off$).  Collisional ionization, for a temperature
that produces {\CIV}, yields a $b ({\rm H})$ too broad to fit the
{\Lya} profile and is, thus, ruled out. For photoionization, the
ionization parameter is constrained by {\NV} to be $-2.0 \la \log U
\la -1.6$. For $\log U \simeq -2.0$, the metallicity is constrained by
the blue wing of the {\Lya} profile to be $\log Z \simeq -1.0$. For
$\log U \simeq -1.6$, the metallicity is $\log Z \simeq -1.2$. The
cloud has a size of $\sim 5$~kpc. Various physical parameters of the
multiple phases in this system are listed in Table~\ref{tab:tab5}.

\subsection{Effects of Assumed Input Spectrum}
\label{sec:specshape}

For the photoionization models, the different spectral energy
distributions (shape) of the ionizing spectrum can produce different
ionization balances in the gas.  Here, we report on our
experimentation with other plausible shapes of the ionizing spectrum.

However, we note that for any one of the systems to be affected by
stellar radiation from one of the galaxies at a galactocentric
distance of $>38 h^{-1}$~kpc would require extreme conditions.  For
the most active bursting galaxy it is reasonable to assume a
luminosity of ionizing photons, $L \sim 10^{54}~{\rm photons~s}^{-1}$,
of which $\sim 1$\% escape \citep{burst}.  At a distance of $38
h^{-1}$~kpc, this corresponds to $5\times10^4~{\rm
photons~cm}^{-2}~{\rm s}^{-1}$.  This is a factor of four less than
the assumed Haardt \& Madau background ionizing photon flux at
$z\sim1$ ($2\times 10^5~{\rm photons~cm}^{-2}~{\rm s}^{-1}$).

The {\OII}~$\lambda$3727 emission flux measured by \citet{thimm95} for
galaxy G2 can be used to estimate the number of ionizing photons
escaping that galaxy.  Any other galaxy, not detected in Thimm's image
would make a smaller contribution.  Following the method of
\citet{kp6}, in \citet{q1206} we estimated that the luminosity of
ionizing photons in G2 is $L \sim 3 \times10^{53}~{\rm
photons~s}^{-1}$.  Although, the escape fraction is highly uncertain,
we note that this luminosity is a factor of three less than that of
the extreme starburst mentioned above.  If an escape fraction of
$1$\% is again assumed, this means the Thimm limit yields an ionizing
photon flux a factor of twelve less than the Haardt \& Madau
background at $1$ Rydberg.

We therefore find it unlikely that systems are {\it significantly\/}
affected by local stellar radiation.  Nonetheless, we briefly consider
the general effects that changes in the shape of our assumed ionizing
spectrum would have on the results above.  Our results here agree with
those in Appendix B of our previous paper on the PG~$1206+459$
absorbers \citep{q1206}

Starburst models with population ages of $0.01$ and $0.1$~Gyrs provide
examples of the most extreme changes in spectral shape we might
expect.  We utilize the burst models of \citet{bruzual}, superimposed on
the $z=1$ Haardt and Madau spectrum, and reconsider the results of our
photoionization models.  Even though they appear to be excluded based
on the \citet{thimm95} limits, we considered models with the stellar flux
equal to the Haardt and Madau flux at $1$ Rydberg.

In a $0.01$~Gyr starburst model there are extreme edges due to {\HI},
{\HeI}, and {\HeII}, leading to a soft spectrum as compared to Haardt
and Madau.  It is clearly impossible to produce significant amounts of
{\CIV}, {\NV}, and {\OVI} absorption by photoionization with this
spectral shape.  Therefore, the extragalactic background photons must
be responsible for the high--ionization phases of the systems, unless
it is collisional ionization.  The low--ionization phase constraints
would be slightly altered by the $0.01$~Gyr starburst spectral shape.
Specifically, the metallicity constraint is typically lowered by
$0.5$~dex.  Furthermore, the {\SiIV} could no longer be produced
primarily in this phase and additional intermediate--ionization phases
would be required.  If subject to the same starburst spectral shape,
it would be difficult to tune these intermediate--ionization phases so
they would not give rise to excess low--ionization absorption.

The $0.1$~Gyr burst model has an extreme {\HI} edge, but above
$1$~Rydberg its shape is similar to that of the Haardt and Madau
spectrum.  Therefore, spectral shape differences produce small changes
in the constraints on photoionized clouds as compared to the pure
Haardt and Madau model.  Slightly more {\SiIV} and {\CIV} would be
produced in the low--ionization phase, which would lead to slightly
lower ionization parameter constraints.  The model metallicity
would actually be increased because the spectral shape of the burst
model provides relatively more photons at energies just above the
ionization edge of {\HI} (where the cross section is largest).

We argued that it is unlikely that anything but small stellar
contributions to the ionizing flux are allowed.  Having explored the
effect of the resulting changes to spectral shape, we conclude that,
even if there are stellar contributions to the ionizing spectrum, this
will not change our basic conclusions.


\section{Summary and Discussion}
\label{sec:discussion}

Along the line of sight toward the quasar PG~$1206+459$ at
$z\sim0.928$ there is a ``group'' of three {\MgII} absorption systems
within a velocity range of $1500$~{\kms}.  With the E230M grating of
STIS on {\it HST}, we obtained $R=30,000$ spectra covering {\Lya} and
various metal line transitions, most notably the high--ionization
transitions of {\SiIV}, {\CIV}, and {\NV}.  These complement our
previous $R=45,000$ Keck/HIRES spectra that showed low--ionization
clouds in {\MgII} and {\FeII}.  We assume that these clouds are slabs,
photoionized by the extragalactic background \citep{haardtmadau96},
and constrain their physical conditions.  We consider any role that
collisional ionization may play in the observed absorption properties.
To account for the entire ensemble of chemical transitions in each
system requires several phases of gas, each with different densities.
We also analyzed a WIYN image of the quasar field and found four
galaxies within $15$\arcsec of the line of sight. The redshift of one
of the galaxies was obtained from long--slit spectra from the KPNO
4--m telescope.

In \S~\ref{sec:discussA}--\ref{sec:discussC}, we will first summarize
the physical conditions that we have found for each system.  We will
also compare to our previous results based on the low--resolution FOS
spectra \citep{q1206}, and to other absorption systems studied at high
resolution.  In \S~\ref{sec:acd}, we will analyze the high--ionization
phases of these systems using methods similar to those applied in the
study of Milky Way high--ionization gas.  The goal of that analysis is
to make comparisons with the Milky Way corona and to assess whether
simple photoionization models can realistically give rise to the
high--ionization absorption.  Finally, in \S~\ref{sec:coso6}, in the
context of our photoionization and collisional ionization models, we
present predictions for the appearance of the {\OVI} profiles that
could be observed at high resolution with Cosmic Origin Spectrograph
(COS), to be commissioned in the 4th servicing mission of {\it HST}.

\subsection{System~A}
\label{sec:discussA}

System A, if it were separated from system B, would be classified as a
multiple--cloud, weak {\MgII} absorber with $W_r(2796) \sim
0.22\pm0.02$~{\AA} \citep{weak1,cv01}.  The six {\MgII} components in
system A are consistent with being produced by $5$--$20$~pc clouds
with $-2.8 \la \log U \la -2.5$.  The metallicity of these clouds
would be $\log Z \simeq 0.5$, with the assumption of a solar abundance
pattern.  The {\NV} profiles are strong and highly structured.  Both
photoionization and collisional ionization models can consistently fit
the profiles, but collisional ionization models require two finely
tuned high--ionization phases. We therefore favor photoionization for
the diffuse phase.  Photoionized regions with sizes of $1$--$65$~kpc,
and with $-1.5 \la \log U \la -0.6$, could produce the {\NV} as well
as the strong {\CIV} and {\OVI} absorption.  This system does not
contribute significantly to the Lyman limit break, again
distinguishing it from strong {\MgII} absorbers \citep{archive1}.

Based on the low--resolution FOS spectrum, \citet{q1206} were unable
to determine whether the high--ionization gas is consistent with
photoionization or collisional ionization.  We have now concluded that
either photoionization or collisional ionization can be consistent
with the relatively narrow {\NV} components. However, a single diffuse
photoionization model producing all three high--ionization transitions
{\CIV}, {\NV}, and {\OVI} is favored.  Our conclusions about the
low--ionization phase are similar to those of our previous study,
except that we infer a slightly lower ionization parameter ($-2.8 \la
\log U \la -2.5$ as compared to $\log U \sim -2.5$).  This is because
we have determined that the {\SiIV}~$\lambda$1394 profile, which is
relatively strong in the FOS spectrum, is contaminated by
{\CIV}~$\lambda$1548 at $z=0.7338$.  Therefore, we now use the
unblended, weaker {\SiIV}~$\lambda$1403 transition in the STIS
spectrum as our constraint.

\begin{figure*}
\figurenum{4} 
\plotone{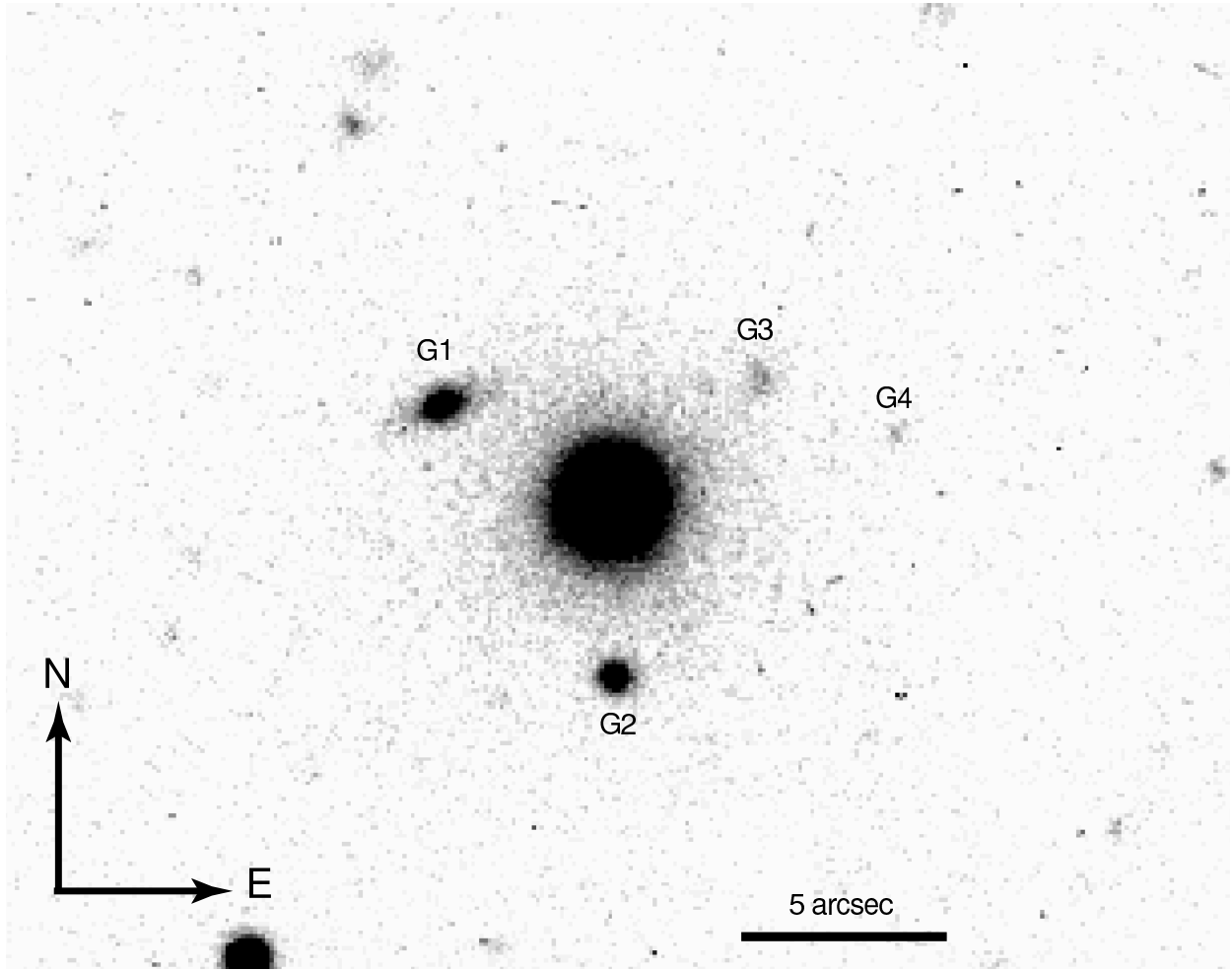} 
\vglue 0.5in
\protect\caption
{The WIYN image of the quasar PG~$1206+459$ field is shown above.  The
bright, central object is the quasar. Four galaxies, G1, G2, G3, and
G4, are detected in the field. G2 has a detected {\OII}~$\lambda$3727
emission line in its long--slit spectrum. G3 is marginally detected in
a Fabry--Perot image, tuned to the redshifted {\OII}~$\lambda$3727
line \citep{thimm95}.}
\label{fig:image}
\end{figure*}

\begin{figure*}
\figurenum{5}
\plotone{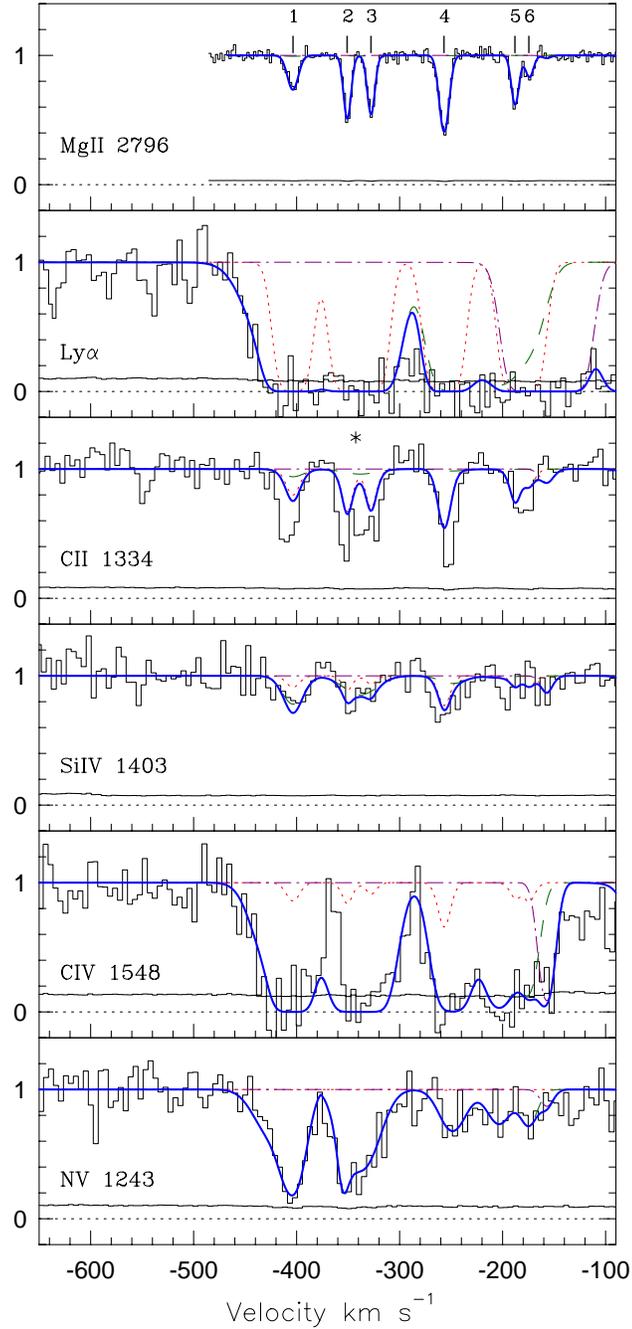}
\vglue -0.5in
\protect\caption
{\footnotesize System A ($z=0.9254$) spectra are shown in the
figure above, with model curves superimposed.  The model parameters
for this system are given in Table~\ref{tab:tab3}.  The histograms,
vertical ticks, and asterisks are as indicated in the caption of
Figure~\ref{fig:data1}. The dotted lines represent contributions from
our model of the {\MgII} clouds. The dashed curves show the
contributions from the photoionized, diffuse phase. The dashed--dotted
lines represent the intermediate, photoionized component. The thick,
solid curve shows the contribution from all phases, which also
includes blended contributions from all three systems.}
\label{fig:fitA}
\end{figure*}

\begin{figure*}
\figurenum{6}
\plotone{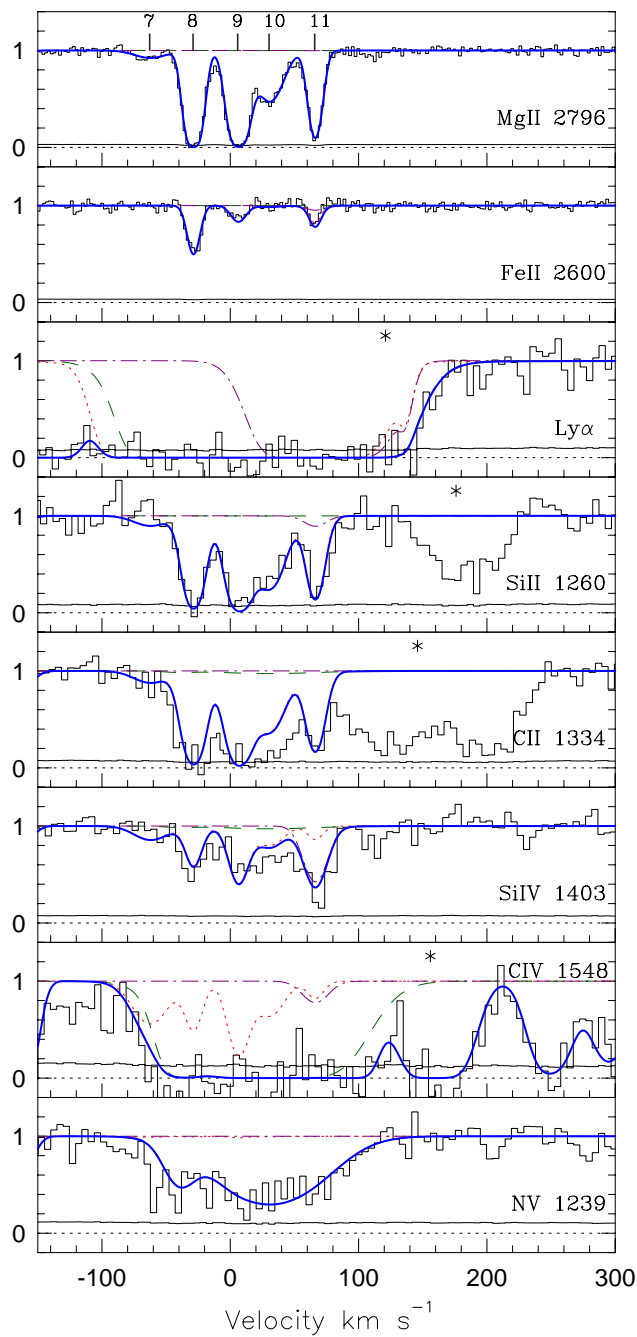}
\vglue -0.5in
\protect\caption
{System B ($z=0.9276$) spectra are shown, with model curves
superimposed.  The model parameters for this system are given in
Table~\ref{tab:tab4}.  The histograms, vertical ticks, and asterisks
are as indicated in the caption of Figure~\ref{fig:data1}.  Model
curves represent the different phases of gas, as outlined in the
caption to Figure~\ref{fig:fitA}.}
\label{fig:fitB}
\end{figure*}

\begin{figure*}
\figurenum{7}
\plotone{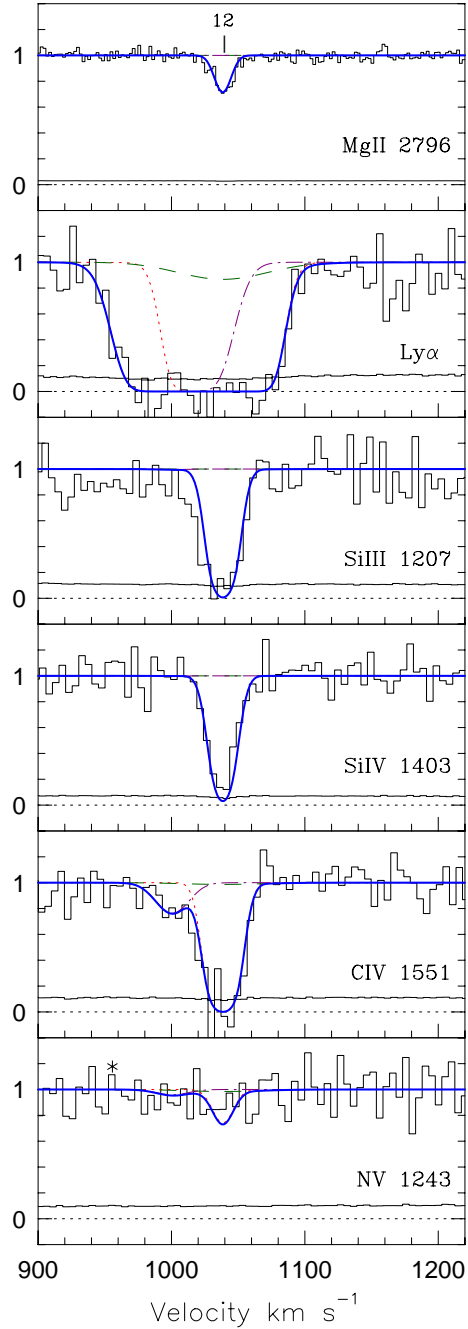}
\vglue -0.5in
\protect\caption
{System C ($z=0.9342$) spectra are shown, with model curves
superimposed.  The model parameters for this system are given in
Table~\ref{tab:tab5}.  The different curves denote separate model
phases as described in the caption to Figure~\ref{fig:fitA}.}
\label{fig:fitC}
\end{figure*}

System A could be associated with one of the galaxies detected in
our WIYN image of the quasar field (Figure~\ref{fig:image}). G2 is
believed to be associated with system B (see \S~\ref{sec:discussB})
and G4 is at a much larger impact parameter than the other three
galaxies. Therefore, G1 and G3, both at an impact parameter of $43
h^{-1}$~kpc, are the most likely host galaxy candidates. They are just
beyond the boundary of $38 h^{-1} (L_K/L_K^*)^{0.15}$~kpc, within
which strong {\MgII} absorption is commonly detected
\citep{bb91,bergeron92,lebrun93,sdp94,s95,3c336}.
G1 appears to be a $\sim 2L^*$ spiral galaxy with a warped disk structure
and G3 is less luminous ($\sim 0.2 L^*$), without apparent morphology.

Both in terms of its {\MgII} equivalent width and the impact parameter
of its candidate host galaxies, system A is just below the threshold
for strong {\MgII} absorption.  It is reasonable to infer that such a
system could arise either in a dwarf galaxy or in the outskirts of a
giant galaxy.  Because the metallicity is supersolar in this case, the
former possibility seems less likely \citep{mateo}.  We favor that
system A is more likely to arise from a patchy distribution of gas in
the outskirts of the apparently warped, spiral disk, G1 in
Figure~\ref{fig:image}.  The high metallicity and large diffuse phase
absorption suggest a local environment with active star
formation. Tidal debris, relating to interactions between the galaxies
in a group, is one possibility.

Multiple--cloud, weak {\MgII} absorbers may be a varied population in
terms of their origins.  A counter--example to system A is the
multiple--cloud, weak {\MgII} absorber at $z=1.0414$ along the line
of sight toward PG~$1634+706$ \citep{zonak}.  In that case, a low
metallicity of $\log Z \sim -1.5$ and the relative kinematics of the
low-- and high--ionization gas suggest an origin in a pair of dwarf
galaxies, a dwarf superwind, or a lower metallicity outer disk region.
That system also differs from system A in that it produces a partial
Lyman limit break.

The most unusual thing about system A is its very strong
high--ionization absorption and, in particular, the structure in
{\NV}.  If the host galaxy is, in fact, the spiral galaxy, G1, we
might expect some similarities with the high--ionization gas phases
around the Milky Way.  We examined the $\sim 20$ archival {\it HST}
E140M spectra of AGNs and quasars that are suitable for study of the
Galactic {\NV} absorption.  Only five had detected Galactic {\NV}
(H~$1821+643$, NGC~$3783$, 3C~$273$, Mrk~$509$, and NGC~$5548$), and
of these only H~$1821+643$ showed relatively narrow components similar
to those in system A.  The H~$1821+643$ line of sight has been studied
in detail \citep{kp1,savage95,oegerle,tripp01}.  The {\NV} absorption,
at $v = -8$~{\kms}, arises within several kpc of the Sun, quite
possibly from the outer boundary of radio loop III \citep{savage95}.
The relatively small ratio, $N({\CIV})/N({\NV})$, is consistent with
conditions expected for collisionally ionized gas in a cooling
supernova bubble \citep{slavincox92,slavincox93}.  The {\NV}
components in system A are even narrower than this special case,
although they are still consistent with an origin in collisionally
ionized gas.  As discussed in \S~\ref{sec:Adiff}, however, if the
{\NV} is produced by collisional ionization, either the {\CIV} or the
{\OVI} would have to arise in a separate phase.

We conclude that the high--ionization phase of system A differs from
those along lines of sight from our vantage point in the Milky Way.
The complex kinematics of the high--ionization gas in system A is
suggestive of dynamic processes perhaps related to interactions
(e.g. \citet{bowen95}).  To understand the physical environment of
system A, we must search for common elements between its galaxy host
properties and those of the hosts of other similar absorption systems
yet to be discovered.

\subsection{System~B}
\label{sec:discussB}

System B, if isolated from system A, would be a classic, strong
{\MgII} absorber with $W_r(2796) \sim 0.66\pm0.01$~{\AA} \citep{cv01}.
It gives rise to a partial Lyman limit break.  Five blended {\MgII}
clouds, with $-3.2 \la \log U \la -2.5$ and sizes of $30$--$200$~pc,
are constrained by the Lyman limit break to have $\log Z \simeq 0.0$.
The cloud farthest to the red ($\sim +50$ {\kms}) has strong, broader
{\SiIV}, which requires an additional, intermediate--ionization phase
(photoionized with $\log U \simeq -2.5$ or collisionally ionized
with $\log T\sim 4.8$~[K]).
The {\NV} absorption profiles are broad ($b ({\rm N}) \sim
50$~{\kms}), and relatively smooth.  They are consistent with being
produced in a $\sim50$~kpc cloud, with $\log U \simeq -1.6$ and $\log
Z \simeq -0.6$ .  The saturated {\CIV} profiles require an additional
$b ({\rm C}) \sim 13$~{\kms} component, offset to the blue.

Our results for system B agree with our previous conclusions based on
the low--resolution FOS spectra \citep{q1206}.  The properties of both
the low-- and high--ionization phases are very similar.  With the
higher--resolution {\SiIV} profiles, we are now able to diagnose its
origin in an additional phase.  Previously, we were able to account
for the overall strength of {\SiIV} by tuning the ionization parameter of the
{\MgII} clouds.

An $\sim L^*$ galaxy at $z \simeq 0.929$, at an impact parameter of
$38 h^{-1}$~kpc (G2 in Figure~\ref{fig:image}), is believed to be
associated with system B.  The galaxy does not have distinguishable
disk structure, so its morphology is uncertain.

The smooth, broad {\NV} absorption profile from system B is quite
similar to most of the {\NV} profiles observed along Milky Way lines
of sight.  For example, the line of sight toward the inner Galaxy
star, HD~167756, shows a smooth, broad {\NV} profile, and a more
structured {\SiIV} profile \citep{savage94}.  Perhaps more relevant
are the extragalactic lines of sight, NGC~3783, 3C~273, Mrk~509, and
NGC~5548, which show similar (but weaker) {\NV} absorption.  In our
Galaxy, the {\NV} absorption is consistent with arising in an
exponential distribution centered around the disk, with scale height
$3.9\pm1.4$~kpc, referred to as the ``Galactic corona'' \citep{MW}.

Because of the similarities with the Milky Way, it is tempting to
interpret the high--ionization gas from system B as a corona.  Also,
the detection of an emission line in the host galaxy of system B is
evidence for star formation activity.  However, there are several
problems with this interpretation.  First, there is no apparent disk
structure in the host galaxy.  Second, the impact parameter is large,
and little is known about corona structure at such distances.  Third,
the {\NV} absorption is extremely strong in this case.  If broad,
smooth, high--ionization components are characteristic of coronae then
we will be likely to find them in many other strong {\MgII} absorbers.
Then, the question would be why these components are so strong in
system B.  Perhaps this could relate to its apparent location in a
group of galaxies.

The intermediate--ionization component needed to explain the {\SiIV}
absorption deserves further comment.  It appears that this type of
situation, where a phase produces absorption only in one transition,
is not uncommon.  The $z=0.9902$ absorber toward PG~$1634+706$
requires an additional collisionally ionized component, with $T\sim
60,000$~K to fit its {\SiIV} profiles \citep{ding}.  Similarly, the
{\SiIII} profiles in the $z=1.0414$ system toward PG~$1634+706$ are
consistent with an additional collisional ionization phase, with
$T\sim 40,000$~K \citep{zonak}.  Cooling proceeds rapidly in this
temperature range, so if these types of gas phases are common it would
imply that formation of shock--heated regions is frequent in
interstellar gas at $z\sim1$.

\subsection{System~C}
\label{sec:discussC}

System C was classified as a single--cloud, weak {\MgII} absorber,
with $W_r(2796) =0.051\pm0.005$~{\AA} \citep{cv01}. Our favored model
for system C includes two clouds at $v \sim +1042$~{\kms} and one at $v
\sim +1005$~{\kms}. The {\MgII} arises in a $\log U \simeq -2.0$ cloud,
which gives rise to most of the observed {\SiIV} and {\CIV}. The
metallicity of this cloud is approximately solar. However, in order to
produce both {\SiII}~$\lambda$1260 and {\SiIV}~$\lambda$1394, a
$0.4$~dex enhancement of silicon is required. Additionally, a
$0.4$~dex reduction of nitrogen is needed to bring the model
prediction into agreement with the data. In our favored model, the
bulk of the {\OVI} arises in a more highly ionized phase, whose
properties are not well constrained because we lack high--resolution
coverage of {\OVI}. The cloud at $v \sim 1005$~{\kms} has an
ionization parameter $\log U \simeq -1.6$ and a metallicity of $\log Z
\sim -1.2$. It is required to produce the blue wing of {\Lya} as well
as the asymmetry in {\CIV}.

In our previous work \citep{q1206}, based on the low--resolution FOS
spectra, we concluded that the {\CIV} absorption could not all be
produced in the same phase as the {\MgII}. However, we could not
distinguish whether the two transitions were aligned in velocity
space.  Now, we see that an offset {\CIV} cloud is needed to explain
the asymmetric {\CIV} profiles. The offset cloud led to a broad {\CIV}
profile in the low--resolution data, and this was the main reason we
found a one--phase model to be inconsistent with the low--resolution
data. However, for more subtle reasons, in our present analysis we
also favor a model with two separate phases centered on the {\MgII}.

Since several galaxies were found at relatively small impact
parameter, it is likely that system C is also related to one of these
galaxies, either G1 or G3.  Since we argued in \S~\ref{sec:discussA}
that G1 was more likely to host system A, by process of elimination we
consider G3 the more likely host for system C.  With a luminosity of
$\sim 0.2 L^*$ and an impact parameter of $43 h^{-1}$~kpc,
statistically, this galaxy would be unlikely to produce strong {\MgII}
absorption.  However, either G3 itself or a high--velocity cloud or
satellite related to G3, could be responsible for this weaker {\MgII}
system.  The average value of the {\OVI} column density in the Small
Magellanic Cloud is $1.7$ times higher than the average value in the
Milky Way \citep{hoopes02}.  The relatively large, high--ionization
absorption from system C could plausibly come from a similar satellite
galaxy.

Single--cloud, weak {\MgII} absorption appears to arise in a variety
of environments.  Clouds with $W_r(2796) \sim 0.1$~{\AA} are found,
both isolated in velocity space \citep{weak1} and clustered within
hundreds of {\kms} of strong {\MgII} absorbers \citep{cv01}.  System C
is an extreme case in that it is a full $1000$~{\kms} from system B.
Single--cloud, weak {\MgII} absorption is also detected in some Milky
Way high--velocity clouds \citep{cwcconf}.  At least some of the
isolated, single--cloud weak {\MgII} absorbers appears to trace tiny
star--forming pockets separate from $L^*$ galaxies \citep{weak2}.
They could arise in faded dwarf galaxies or in remnants of Population
III star clusters.  The nature of the high--velocity clouds also
remains a mystery: are they ejected from the Milky Way disk, or are
they primordial clouds falling into the Local Group? (see, e.g.,
\citet{wakker97}, \citet{blitz02}, and \citet{putman}).  More
generally, the question is whether the various weak {\MgII} absorption
profiles are produced by the same physical structures in different
environments, or whether the similarity of their {\MgII} profiles is
just a coincidence.

We compare system C to the three isolated, single--cloud, weak {\MgII}
absorbers along the line of sight toward PG~$1634+706$ \citep{ding}.
All three of them have close to solar metallicity, similar to system
C, and two of them require additional high--ionization clouds, offset
in velocity.  However, they differ from system C in that they all
require a lower--ionization phase in which the {\MgII} absorption
arises (with $-4.5 \la \log U \la -2.7$) and a separate
high--ionization phase (with $-2.5 \la \log U \la -1.5$) to produce
the {\CIV} absorption centered on the {\MgII}.  It is interesting that
the phase that produces {\MgII} absorption for system C (with $\log U
\sim -2.0$) is more similar to the high--ionization phase for the
three isolated, single--cloud, weak {\MgII} absorbers.  The size of
the {\MgII} cloud in system C is $4$~kpc, while the low--ionization
clouds that produce isolated, single--cloud weak {\MgII} absorbers are
quite small (pcs or tens of pcs).

Another interesting comparison is to the individual {\MgII} clouds in
system A, since they each have about the same equivalent width as
system C.  There are small differences in the inferred properties of
the phases giving rise to {\MgII}.  System A clouds generally have a
slightly lower $U$, a slightly higher (supersolar) metallicity, and
smaller $b$ parameters.  However, the most striking difference is
between the high--ionization diffuse phases.  The separate, diffuse
clouds of system A have very strong {\NV} absorption, but {\NV} is not
detected in system C.  In fact, system C is the more typical absorber,
since the {\NV} absorption is unusually strong in system A (and system
B).

There is presently little direct information published about the
physical conditions of the gas in ``satellite clouds'' of strong
{\MgII} absorbers at $z\sim1$ which are found within hundreds of
{\kms} of the dominant absorption component.  It is likely that they
are sub--Lyman limit systems \citep{cv01}, and many have a resolved
{\CIV} component separate from that related to the dominant,
low--ionization component (based upon our unpublished STIS data).
Some also have detected {\OVI}.  In principal, system C could simply
be a similar satellite cloud, though at a much larger velocity
separation than the others.

A related comparison is to the high--velocity clouds around the Milky
Way, not those that are detected by $21$--cm emission, but those that
are below that threshold and are detected through absorption in
various metal transitions.  The Galactic HVCs are a varied population
in terms of the phase structure, with some having strong
low--ionization absorption and others having only high--ionization
absorption \citep{sembachhvc}.  Some systematic differences between
Milky Way HVCs and those at $z\sim1$ are to be expected because the
extragalactic background contribution has decreased substantially
since $z\sim1$ \citep{haardtmadau96}.  In particular, some Milky Way
HVCs are detected only in low--ionization transitions, which could be
explained by a stellar contribution to the ionizing spectrum. However,
the ``satellite clouds'' around strong {\MgII} absorbers at $z\sim1$
and the HVCs around the Milky Way could really be the same population
\citep{cv01,cwcconf}.

In conclusion, the physical conditions we inferred for system C could
be consistent with it arising in the same type of structure that
produces satellite absorption at smaller velocities around various
strong {\MgII} absorbers.  These conditions are not consistent with
the properties of isolated, weak {\MgII} absorbers
\citep{weak2,weak1634}.  In order that the probability of intercepting
a high--velocity cloud at $+1000$~{\kms} is not vanishingly small
(e.g. among the 22 high--velocity clouds in \citet{cv01}, the 2 with
the highest velocity are at $\sim \pm410$~{\kms}), there would have to
be an abundant population of these objects in the PG~$1206+459$
$z\sim0.93$ group, and their cross section would have to be large.

\subsection{Nature of the High--Ionization Phases}
\label{sec:acd}

Because of inefficiencies of observing in the ultraviolet, there are
very few examples of high--resolution quasar absorption line systems
at $z \sim 1$ with high--resolution spectra covering many transitions.
By far, the most extensive body of detailed work analyzing spectra
probing interstellar medium and halo gas comes from studies of our
Galaxy.  \citet{MW} considered the mechanisms responsible for
producing and maintaining the Galactic corona.  In particular, they
determined the apparent column density ratios of {\SiIV}, {\CIV}, and
{\NV} at various scale--heights from the disk, and in various
directions.  They argued that a simple photoionization or collisional
ionization model could not reproduce the range of observed values, and
compared their observations to results from models of more complex,
non--equilibrium processes.  The conclusion of their work was that a
hybrid model that includes some combination of several such processes
(e.g., Galactic fountain gas, gas related to supernova bubbles, and
turbulent mixing layers of hot and warm gas) is needed.

\begin{figure*}
\figurenum{8}
\plotone{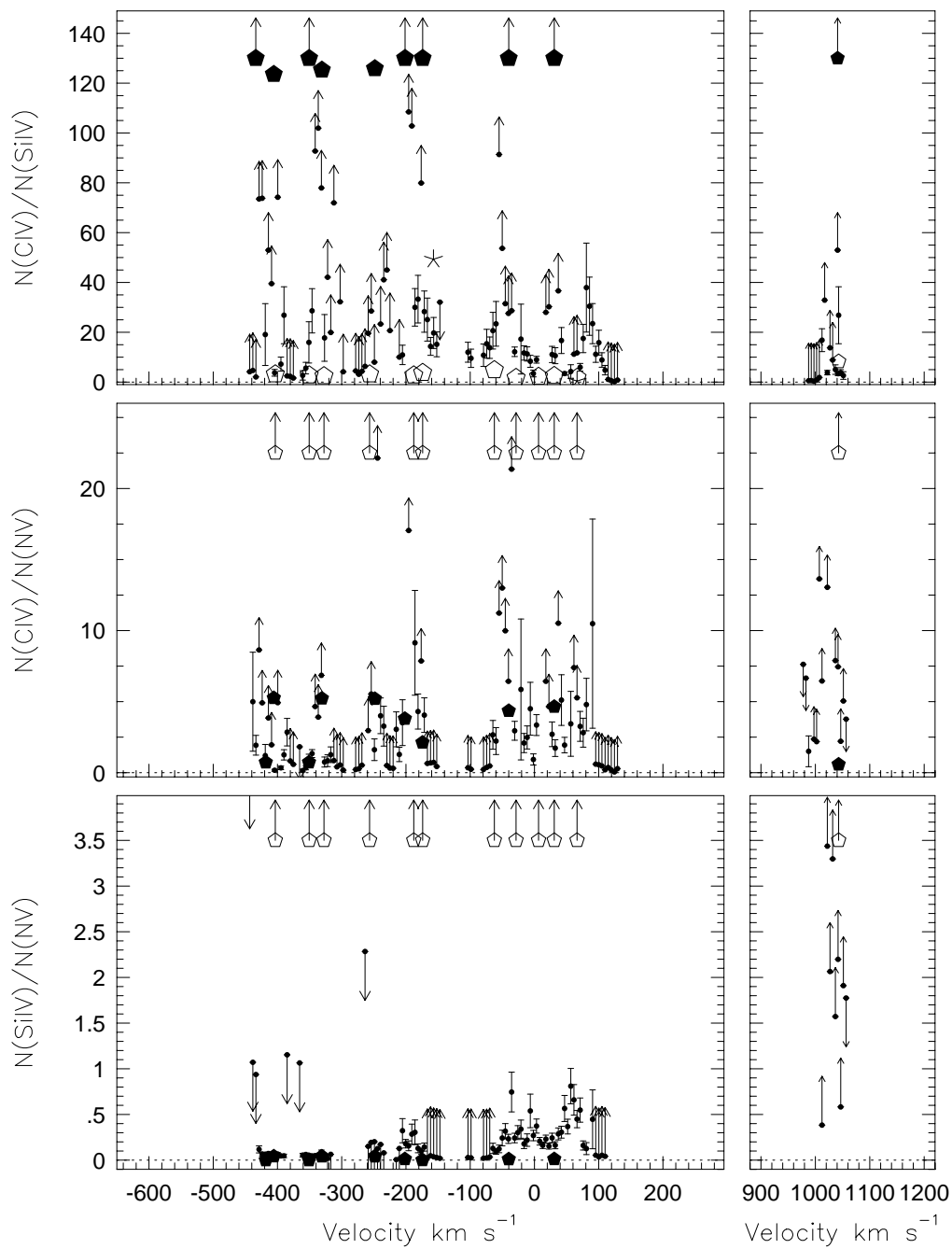}
\vglue -0.5in
\protect\caption
{\scriptsize
The ratios of apparent column density between {\SiIV}, {\CIV}, and
{\NV} are displayed for all three systems in velocity space above. The
small dots represent the unsaturated {\it HST}/STIS data, with $1\sigma$
errors superimposed. The upper/lower data limits are denoted by small
dots with upper/lower arrows superimposed. The open pentagons
represent the model values for our low--ionization clouds, the filled
pentagons show the contribution from the diffuse phase, and the
skeletal point represents the intermediate component. The pentagons
with upper arrows superimposed indicate that the actual ratios are too
large to be displayed on the plots.}
\label{fig:ratio}
\end{figure*}

We can compare the properties of the high--ionization gas phases of
systems A, B, and C to those of the gas around the Galaxy.  To
facilitate this comparison, we calculated the apparent column
densities of {\SiIV}, {\CIV}, and {\NV} as a function of velocity
along the PG~$1206+459$ line of sight.  The apparent column density of
a given transition at velocity $v$ is given (as in \citet{acd}) by
$$
N_a(v) [{\rm cm}^{-2} ({\kms})^{-1}] = (m_e c/\pi e^2) (f
\lambda)^{-1} \tau_a (v),
$$
where $f$ is the oscillator strength of
the transition and $\lambda$ is in {\AA}.  The apparent optical depth,
$\tau_a (v)$ is given by 
$$
\tau_a (v) = \ln [1/I_{norm}(v)],
$$ where $I_{norm}$ is the normalized flux.  Values of $N_a(v)$ were
obtained in every pixel in which a transition was detected at a
$3\sigma$ level. Otherwise, limits were derived by replacing
$I_{norm}(v)$ by $1-\sigma(v)$, where $\sigma(v)$ is the $1\sigma$
error bar on the flux. We computed $N_a(v)$ for {\SiIVdblt},
{\CIVdblt}, and {\NVdblt}.  For each doublet, a single apparent column
density value was derived based on the following rules: 1) if neither
transition was saturated, the $N_a(v)$ values were averaged; 2) if
only the stronger member of the doublet was saturated we used the
weaker member; and 3) if both members of the doublet were saturated,
we used the weaker member to set a lower limit.  Measurements or
limits for the ratios $N_a({\CIV})/N_a({\SiIV})$,
$N_a({\CIV})/N_a({\NV})$, and $N_a({\SiIV})/N({\NV})$ were determined
in regions where at least one transition was detected.  These ratios,
with $1\sigma$ errors or limits where appropriate, are shown as a
function of velocity, in Figure~\ref{fig:ratio}.

The apparent column density ratios for systems A, B, and C can be
directly compared to those found in our Galaxy (Figure 6 of
\citet{MW}).  In general, systems A and B have
$N_a({\CIV})/N_a({\SiIV})$ much larger and $N_a({\SiIV})/N_a({\NV})$
much smaller than any Milky Way line of sight.  Taken at face value,
all three ratios are consistent only with the ``interface/bubble''
model and not with ``fountains'' or with ``mixing layers'' (for
details and references on the different models, see \citet{MW}).  In
system C, {\NV} is not detected, but the lower limits on
$N_a({\SiIV})/N_a({\NV})$ and $N_a({\CIV})/N_a({\NV})$, and the large
value of $N_a({\CIV})/N_a({\SiIV})$ are consistent only with the
``mixing layers'' model.

We need to be cautious about this simple interpretation.
Figure~\ref{fig:ratio} also shows, as solid pentagons, the ratios of
the cloud column densities for the high--ionization phase of our
model.  In our models for all three systems, {\SiIV} is found
primarily in the low--ionization component, so
$N_a({\CIV})/N_a({\SiIV})$ is off scale for the high--ionization
components.  For the low--ionization component,
$N_a({\CIV})/N_a({\SiIV}) \sim 2$ for systems A and B.  The observed
values are actually in between those of the two model phases. This is
not unexpected, since we are averaging along a line of sight that is
likely to pass through many types of regions.  It is interesting to
note that, even for our Galaxy, the {\OVI} column density varies by
factors of several on scales of hundreds of parsecs \citep{howk02}.
Since the high--ionization gas has such a patchy distribution, we
would expect the local apparent column density ratios to vary
significantly from the global value.

We conclude, that if the gas is in a complex set of phases along an
extended line of sight, as we have proposed, it is inappropriate to
directly compare the observed apparent column density ratios to any
models.  For the Milky Way, the interpretation is aided by the fact
that specific regions can be isolated in velocity space, and ``local''
apparent column density ratios can be considered.  For extragalactic
absorption systems, understanding the phase structure is critical to
isolating the regions, and therefore to interpreting what processes
are involved in producing and maintaining the high--ionization gas.

\subsection{Expectations for High--Resolution {\OVI} Profiles}
\label{sec:coso6}

It is a bit surprising that simple photoionization models can
consistently produce the absorption observed in all of the
high--ionization transitions.  This is especially true, in view of the
fact that such simple models fail to explain the Milky Way corona
\citep{MW}.  Fortunately, a test of the simple models is possible.

\begin{figure*}
\figurenum{9} 
\plotone{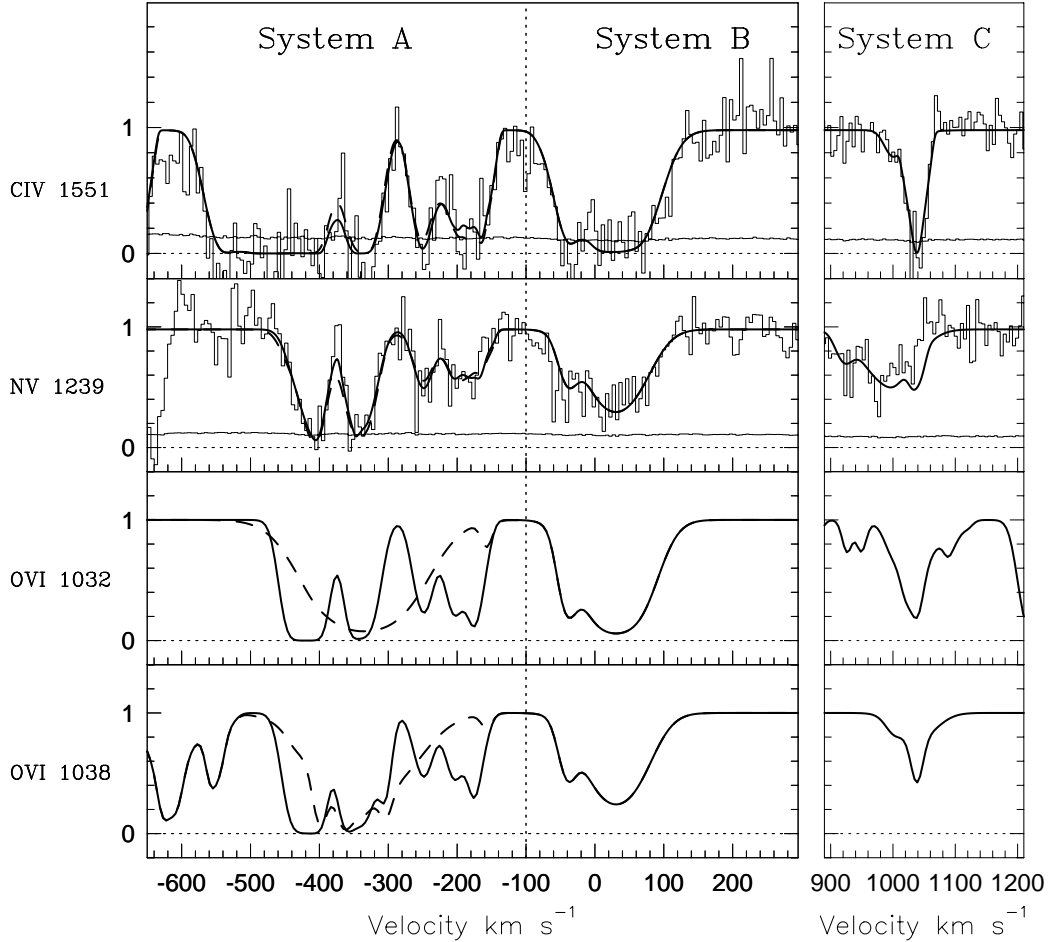} 
\vglue -2.5in 
\protect\caption 
{Simulated models of {\OVIdblt} as it would be observed by the {\it
HST}/COS ($R=18,000$) are displayed above as solid curves.  The
simulations use the model parameters listed in
Table~\ref{tab:tab3}--\ref{tab:tab5}, with a photoionized component in
the diffuse phase of system C.  Solid curves are for the
photoionization model for system A, and dashed lines for the
collisional ionization model.  The {\CIV} and {\NV} profiles are also
displayed for comparison.  The synthetic model curves are superimposed
on the STIS data covering these two transitions.}
\label{fig:cos}
\end{figure*}

We have been able to place good constraints on the properties of the
photoionized high--ionization phases based particularly on the
{\SiIV}, {\CIV}, and {\NV} transitions, as well as on the
low--resolution FOS profile of {\OVIdblt}.  Though both have extremely
strong absorption in high--ionization transitions, systems A and B are
strikingly different.  System A appears to be the unusual one because
of the relatively narrow {\NV} structures.  High resolution coverage
of the {\OVI} would be essential to distinguish between collisional
ionization (which is permitted by the data) and photoionization (the
minimal phase model which we prefer).  System B is consistent with
what one would expect from corona gas, with broad, and less
structured, high--ionization profiles.  The model constraints for
these two systems provide a very specific prediction for how the
{\OVIdblt} would look at higher resolution.  Figure~\ref{fig:cos}
presents the model predictions for an observation of {\OVIdblt} using
the planned COS on {\it HST}, which will have a resolution of $R =
18,000$ at the observed wavelength of {\OVI}.  The solid curve
represents a pure photoionization model for all three systems, as
summarized in Tables~\ref{tab:tab3}--\ref{tab:tab5}.  The dashed curve
indicates the prediction of a collisional ionization model for system
A.  We find that if a simple, single--phase photoionization model
applies for system A, the {\OVI} should have structure very similar to
the {\NV}.  If a collisional ionization model applies, the {\OVI}
profiles for system A would be smooth, similar to those of system B.
For the case of system C, there are a wide range of parameters allowed
for photoionization and collisional ionization production of {\OVI},
and only one model prediction is shown in Figure~\ref{fig:cos}.
Higher--resolution {\OVI} data would be needed to distinguish between
photoionization and collisional ionization of the phase of gas that
gives rise to the {\OVI} absorption in system C.

Higher--resolution coverage of {\OVI} for the absorption systems at
$z\sim 0.93$ is important to test the validity of the modeling
techniques we have used here, and to better understand the physical
conditions in the gas.  However, we also note that the basic
conclusions of the present work, with our new STIS/E230M observations,
are not substantially different than those that we reached based upon
lower--resolution FOS spectra, using Keck/HIRES observations of
{\MgII} as a ``template''.  Therefore, we probably already have a
reasonably accurate view of the physical conditions in the gas.  What
is really needed to interpret this set of absorption systems along the
PG~$1206+459$ line of sight, is to place them into the context of a
larger, representative sample. It is already apparent that more quasar
lines of sight need to be observed in order to fully sample the
variety of galaxy morphologies, environments, and local processes that
influence the gas.

\acknowledgements
Support for this work was provided by the NSF (AST 96--17185), and by
NASA (NAG 5--6399 and STSI GO--08672.01--A).
We thank Buell Jannuzi, Karen Kneirman, and Rajib Ganguly for their assistance with and 
participation in the CryoCam observations.  We also thank Buell Jannuzi for 
helpful comments on this manuscript.  Blair Savage and Todd Tripp provided
useful insights that helped us to refine our results.  Finally, we thank the
staff at the Space  Telescope Science Institute, especially Tricia Royle, for their 
excellent services.


\clearpage

\clearpage

\clearpage

\clearpage

\clearpage

\clearpage

\clearpage

\clearpage

\clearpage

\newpage

\begin{deluxetable}{cccc}
\tablenum{1}

\tabletypesize{\footnotesize}
\rotate
\tablewidth{0pt}
\tablecaption{Rest Frame Equivalent Widths}
\tablehead{
\colhead{} &
\colhead{System A}  &
\colhead{System B}  &
\colhead{System C} \\
\colhead{} &
\colhead{$z=0.92540$} &
\colhead{$z=0.92760$} &
\colhead{$z=0.93428$}
}

\startdata
\hline
{\Lya}        & $<2.38\pm0.03$    & $<2.38\pm0.03$ & $0.55\pm0.02$   \\
{\MgI}~$\lambda$2853   & $<0.006$          & $0.04\pm0.02$  & $<0.006$        \\
{\MgII}~$\lambda$2796  & $0.22\pm0.0.02$ & $0.66\pm0.01$  & $0.05\pm0.01$ \\
{\MgII}~$\lambda$2803  & $0.12\pm0.0.02$ & $0.51\pm0.01$  & $0.02\pm0.01$ \\ 
{\FeII}~$\lambda$2600  & $<0.006$          & $0.08\pm0.02$  & $<0.006$        \\
{\SiII}~$\lambda$1260  & $0.16\pm0.03$     & $0.34\pm0.02$    & $0.16\pm0.01$ \\
{\CII}~$\lambda$1335   & $<0.25\pm0.03$    & $0.94\pm0.02$   & $0.10\pm0.01$  \\
{\SiIII}~$\lambda$1207 & $0.78\pm0.09$     & $0.66\pm0.02$   & $0.14\pm0.02$  \\
{\SiIV}~$\lambda$1394  & $<0.41\pm0.02$    & $0.46\pm0.01$   & $0.10\pm0.01$   \\
{\SiIV}~$\lambda$1403  & $0.10\pm0.02$     & $0.28\pm0.02$   & $0.11\pm0.01$   \\
{\CIV}~$\lambda$1548   & $1.28\pm0.08$& $1.98\pm0.20$ &$0.31\pm0.02$  \\
{\CIV}~$\lambda$1551   & $1.98\pm0.20$& $0.91\pm0.07$& $0.22\pm0.02$      \\
{\NV}~$\lambda$1239    & $0.76\pm0.06$    & $0.28\pm0.02$    & $<0.02$        \\
{\NV}~$\lambda$1243    & $0.28\pm0.02$    & $0.21\pm0.02$    & $<0.02$      \\
\hline
\enddata
\vglue -0.05in

\tablecomments{
\baselineskip=0.7\baselineskip
 Upper limits for non-detections are at a $3\sigma$ level.
Upper limits with specified error bars are from blends that could not
be distinctly separated by Gaussian fits.}
\label{tab:tab1}
\end{deluxetable}

\newpage

\begin{deluxetable}{cccc}
\tablenum{2}

\tabletypesize{\footnotesize}
\rotate
\tablewidth{0pt}
\tablecaption{Galaxy Properties}
\tablehead{
\colhead{} &
\colhead{luminosity}  &
\colhead{Impact Parameter} &
\colhead{redshift} \\
\colhead{} &
\colhead{[$L_K^*$]} &
\colhead{[$h^{-1}~kpc$]} &
\colhead{}
}

\startdata
G1 & $2$ & $43$ & \nodata \\
\hline

G2 & $1$ & $38$ & $0.9289$ \\
\hline

G3 & $0.2$ & $43$ & $\sim0.93$ \\
\hline

G4 & $0.1$ & $65$ & \nodata \\

\enddata
\tablecomments{The impact parameters are calculated assuming $z \sim 0.93$.}

\label{tab:tab2}

\end{deluxetable}

\begin{deluxetable}{lccccrcccccccccc}
\tablenum{3}

\tabletypesize{\tiny}
\rotate
\tablewidth{0pt}
\tablecaption{System A at $z=0.9254$}
\tablehead{
\colhead{} &
\colhead{$v$}  &
\colhead{$Z$} &
\colhead{$\log U$} &
\colhead{$\log n_H$} &
\colhead{size} &
\colhead{$T$} &
\colhead{$N_{\rm tot}({\rm H})$} &
\colhead{$N({\HI})$} &
\colhead{$N({\MgII})$} &
\colhead{$N({\SiIV})$} &
\colhead{$N({\CIV})$} &
\colhead{$N({\NV})$} &
\colhead{$b({\rm H})$} &
\colhead{$b({\rm Mg})$} &
\colhead{$b({\rm N})$} \\
\colhead{} &
\colhead{[{\kms}]} &
\colhead{[$Z_{\odot}$]} &
\colhead{} &
\colhead{[{\cc}]} &
\colhead{[kpc]} &
\colhead{[K]} &
\colhead{[{\cmsq}]} &
\colhead{[{\cmsq}]} &
\colhead{[{\cmsq}]} &
\colhead{[{\cmsq}]} &
\colhead{[{\cmsq}]} &
\colhead{[{\cmsq}]} &
\colhead{[{\kms}]} &
\colhead{[{\kms}]} &
\colhead{[{\kms}]}
}

\startdata
\multicolumn{16}{c}{\sc {\MgII}~Phase}\\
\hline
{\MgII}$- \rm 1$ & $-403$ & $3$ & $-2.6$ & $-2.6$ & $0.007$ & $2300$ & $16.8$ & $14.7$ & $11.9$ & $12.2$ & $12.5$ & $11.1$ & $10.5$ & $6.0$ & $6.2$\\
{\MgII}$- \rm 2$ & $-351$ & $3$ & $-2.7$ & $-2.5$ & $0.008$ & $2200$ & $16.9$ & $14.9$ & $12.2$ & $12.2$ & $12.5$ & $11.1$ & $8.9$ & $2.9$ & $3.3$\\
{\MgII}$- \rm 3$ &   $-328$ & $3$  & $-2.8$ & $-2.4$ & $0.005$ & $2200$ & $16.8$ & $14.9$ & $12.1$ & $12.0$ & $12.2$ & $10.7$ & $9.0$ & $3.0$ & $3.3$\\
{\MgII}$- \rm 4$ &  $-257$ & $3$ & $-2.6$ & $-2.6$ & $0.02$ & $2300$ & $17.2$ & $15.1$ & $12.4$ & $12.6$ & $12.9$ & $11.5$ & $9.8$ & $4.8$ & $5.0$\\
{\MgII}$- \rm 5$ & $-188$ & $3$ & $-2.7$ & $-2.5$ & $0.006$ & $2200$ & $16.7$ & $14.8$ & $12.0$ & $12.1$ & $12.3$ & $10.9$ & $9.0$ & $3.1$ & $3.4$\\
{\MgII}$- \rm 6$ & $-174$ & $3$ & $-2.5$ & $-2.7$ & $0.006$ & $2500$ & $16.6$ & $14.4$ & $11.6$ & $12.1$ & $12.4$ & $11.1$ & $9.6$ & $3.7$ & $4.0$\\

\hline
\multicolumn{16}{c}{\sc Intermediate~Phase} \\
\hline
{\CIV}$- \rm ph$ & $-157$  & $0.1$ & $-2.0$ & $-3.2$ & $5$ & $12000$ & $17.9$ & $15.6$ & $10.7$ & $12.3$ & $14.0$ & $12.9$ & $19.8$ & $4.4$ & $5.6$\\
\hline
{\CIV}$- \rm co$ & $-157$  & $0.01$ & \nodata & \nodata & \nodata & $100000$ & $20.0$ & $15.2$ & $9.4$ & $12.2$ & $14.0$ & $11.2$ & $4.2$ & $5.8$ & $5.6$\\
\hline\hline

&
$v$ &
$Z$ &
$\log U$ &
$\log n_H$ &
size &
$T$ &
$N_{\rm tot}({\rm H})$ &
$N({\HI})$ &
$N({\OVI})$ &
$N({\SiIV})$ &
$N({\CIV})$ &
$N({\NV})$ &
$b({\rm H})$ &
$b({\rm O})$ &
$b({\rm N})$ \\
 &
[{\kms}] &
[$Z_{\odot}$] &
 &
[{\cc}] &
[kpc] &
[K] &
[{\cmsq}] &
[{\cmsq}] &
[{\cmsq}] &
[{\cmsq}] &
[{\cmsq}] &
[{\cmsq}] &
[{\kms}] &
[{\kms}] &
[{\kms}] \\

\hline
\multicolumn{16}{c}{\sc {Diffuse}~Phase (Case A: Photoionization)} \\
\hline
{\NV}$- \rm 1$ & $-424$  & $3$ & $-0.6$ & $-4.6$ & $31$ & $25000$ & $18.4$ & $13.6$ & $15.1$ & $8.1$ & $13.7$ & $13.8$ & $28.7$ & $21.0$ & $21.1$\\
{\NV}$- \rm 2$ & $-403$   & $3$ & $-1.5$ & $-3.7$ & $3$ & $8000$ & $18.3$ & $14.8$ & $14.7$ & $12.8$ & $14.9$ & $14.2$ & $16.6$ & $12.6$ & $12.6$\\
{\NV}$- \rm 3$ & $-355$  & $3$ & $-0.6$ & $-4.6$ & $65$ & $25000$ & $18.7$ & $13.9$ & $15.4$ & $8.4$ & $14.0$ & $14.2$ & $10.9$ & $2.7$ & $2.9$\\
{\NV}$- \rm 4$ & $-338$ & $3$ & $-1.5$ & $-3.7$ & $3$ & $8000$ & $18.3$ & $14.8$ & $14.7$ & $12.8$ & $14.9$ & $14.2$ & $23.7$ & $21.1$ & $21.1$\\
{\NV}$- \rm 5$ & $-248$ & $3$ & $-1.5$ & $-3.7$ & $1$ & $8000$ & $17.4$ & $14.3$ & $14.1$ & $12.3$ & $14.4$ & $13.7$ & $19.1$ & $15.7$ & $15.7$\\
{\NV}$- \rm 6$ & $-204$ & $3$ & $-1.4$ & $-3.8$ & $1$ & $10000$ & $17.8$ & $13.9$ & $14.1$ & $11.6$ & $14.1$ & $13.5$ & $18.7$ & $14.0$ & $14.1$\\
{\NV}$- \rm 7$ & $-175$ & $3$ & $-1.2$ & $-4.0$ & $1$ & $15000$ & $17.6$ & $13.4$ & $14.2$ & $10.5$ & $13.7$ & $13.4$ & $17.5$ & $9.0$ & $9.1$\\

\hline
\multicolumn{16}{c}{\sc {Diffuse}~Phase (Case B: Collisional Ionization)} \\
\hline
{\NV}$- \rm 1$ & $-418$  & $3$ & \nodata & \nodata & \nodata & $150000$ & $18.0$ & $12.8$ & $11.9$ & $11.8$ & $13.7$ & $13.6$ & $52.7$ & $20.1$ & $21.3$\\
{\NV}$- \rm 2$ & $-406$   & $3$ & \nodata & \nodata & \nodata & $150000$ & $18.8$ & $13.6$ & $12.6$ & $12.6$ & $14.4$ & $14.3$ & $50.5$ & $14.2$ & $15$\\
{\NV}$- \rm 3$ & $-351$  & $3$ & \nodata & \nodata & \nodata & $150000$ & $18.5$ & $13.2$ & $12.3$ & $12.3$ & $14.1$ & $14.0$ & $50.8$ & $15.3$ & $16$\\
{\NV}$- \rm 4$ & $-331$ & $3$ & \nodata & \nodata & \nodata & $140000$ & $18.7$ & $13.6$ & $11.9$ & $12.7$ & $14.6$ & $14.0$ & $48.7$ & $15.4$ & $16$\\
{\NV}$- \rm 5$ & $-248$ & $3$ & \nodata & \nodata & \nodata & $140000$ & $18.3$ & $13.2$ & $11.4$ & $12.3$ & $14.2$ & $13.6$ & $49.0$ & $16.2$ & $16.8$\\
{\NV}$- \rm 6$ & $-202$ & $3$ & \nodata & \nodata & \nodata & $140000$ & $18.3$ & $13.2$ & $11.5$ & $12.2$ & $14.2$ & $13.6$ & $48.8$ & $15.7$ & $16.3$\\
{\NV}$- \rm 7$ & $-174$ & $3$ & \nodata & \nodata & \nodata & $140000$ & $18.1$ & $13.0$ & $11.2$ & $12.0$ & $14.0$ & $13.4$ & $48.4$ & $14.3$ & $15$\\
\hline
{\OVI}$- \rm ph$ & $-335$ & $3$ & $-0.6$ & $-4.6$ & $31$ & $250000$ & $18.4$ & $13.6$ & $15.0$ & $7.92$ & $13.6$ & $13.7$ & $82.4$ & $80.0$ & $80.0$\\
\hline
{\OVI}$- \rm co$ & $-335$ & $3$ & \nodata & \nodata & \nodata & $320000$ & $18.5$ & $12.6$ & $15.0$ & $11.1$ & $12.7$ & $13.1$ & $106$ & $80.0$ & $80.3$\\

\hline
\enddata
\vglue -0.05in

\tablecomments{
\baselineskip=0.7\baselineskip
Column densities are listed in logarithmic units. In the
intermediate--ionization phase, both the photoionization and
the collisional ionization cases are listed.  In Case B for the
diffuse phase, there is a collisionally ionized {\NV} accompanied
by either the photoionized or the collisionally ionized {\OVI}
cloud listed directly below.}

\label{tab:tab3}
\end{deluxetable}

\begin{deluxetable}{lrcclccccccccccc}
\tablenum{4}

\tabletypesize{\tiny}
\rotate
\tablewidth{0pt}
\tablecaption{System B at $z=0.9276$}
\tablehead{
\colhead{} &
\colhead{$v$}  &
\colhead{$Z$} &
\colhead{$\log U$} &
\colhead{$\log n_H$} &
\colhead{size} &
\colhead{$T$} &
\colhead{$N_{\rm tot}({\rm H})$} &
\colhead{$N({\HI})$} &
\colhead{$N({\MgII})$} &
\colhead{$N({\SiIV})$} &
\colhead{$N({\CIV})$} &
\colhead{$N({\NV})$} &
\colhead{$b({\rm H})$} &
\colhead{$b({\rm Mg})$} &
\colhead{$b({\rm N})$} \\
\colhead{} &
\colhead{[{\kms}]} &
\colhead{[$Z_{\odot}$]} &
\colhead{} &
\colhead{[{\cc}]} &
\colhead{[kpc]} &
\colhead{[K]} &
\colhead{[{\cmsq}]} &
\colhead{[{\cmsq}]} &
\colhead{[{\cmsq}]} &
\colhead{[{\cmsq}]} &
\colhead{[{\cmsq}]} &
\colhead{[{\cmsq}]} &
\colhead{[{\kms}]} &
\colhead{[{\kms}]} &
\colhead{[{\kms}]}
}

\startdata
\multicolumn{16}{c}{\sc {\MgII}~Phase}\\
\hline
{\MgII}$- \rm 7$ & $-62$ & $1.0$ & $-2.5$ & $-2.7$ & $0.1$ & $10000$ & $17.7$ & $15.1$ & $11.7$ & $12.7$ & $13.4$ & $11.8$ & $24.5$ & $16.2$ & $16.5$\\
{\MgII}$- \rm 8$ & $-29$ & $1.0$ & $-3.2$ & $-2$ & $0.1$ & $8000$ & $18.6$ & $16.7$ & $13.4$ & $13.0$ & $13.2$ & $11.3$ & $17.3$ & $5.7$ & $6.4$\\
{\MgII}$- \rm 9$ &   $6$ & $1.0$  & $-3.0$ & $-2.2$ & $0.2$ & $9000$ & $18.6$ & $16.5$ & $13.3$ & $13.2$ & $13.6$ & $11.7$ & $18.3$ & $7.4$ & $8.0$\\
{\MgII}$- \rm 10$ &  $31$ & $1.0$ & $-2.9$ & $-2.3$ & $0.1$ & $9000$ & $18.0$ & $15.8$ & $12.6$ & $12.8$ & $13.2$ & $11.4$ & $21.3$ & $12.7$ & $13.1$\\
{\MgII}$- \rm 11$ & $66$ & $1.0$ & $-3.2$ & $-2$ & $0.03$ & $8500$ & $18.0$ & $16.1$ & $12.8$ & $12.4$ & $12.6$ & $10.7$ & $17.4$ & $5.1$ & $5.9$\\

\hline
\multicolumn{16}{c}{\sc Intermediate~Phase} \\
\hline
{\SiIV}$- \rm ph$ & $66$  & $0.25$ & $-2.5$ & $-2.7$ & $2$ & $16000$ & $19.0$ & $16.2$ & $12.3$ & $13.3$ & $14.1$ & $12.5$ & $18.8$ & $10.0$ & $10.5$\\
\hline
{\SiIV}$- \rm co$ & $66$  & $0.25$ & \nodata & \nodata & \nodata & $63000$ & $18.8$ & $14.7$ & $10.0$ & $13.3$ & $12.9$ & $7.6$ & $33$ & $10.3$ & $11.7$\\

\hline\hline
&
$v$ &
$Z$ &
$\log U$ &
$\log n_H$ &
size &
$T$ &
$N_{\rm tot}({\rm H})$ &
$N({\HI})$ &
$N({\OVI})$ &
$N({\SiIV})$ &
$N({\CIV})$ &
$N({\NV})$ &
$b({\rm H})$ &
$b({\rm O})$ &
$b({\rm N})$ \\
 &
[{\kms}] &
[$Z_{\odot}$] &
 &
[{\cc}] &
[kpc] &
[K] &
[{\cmsq}] &
[{\cmsq}] &
[{\cmsq}] &
[{\cmsq}] &
[{\cmsq}] &
[{\cmsq}] &
[{\kms}] &
[{\kms}] &
[{\kms}] \\

\hline
\multicolumn{16}{c}{\sc {Diffuse}~Phase} \\
\hline
{\CIV}$- \rm 8$ & $-40$  & $0.25$ & $-1.6$ & $-3.6$ & $8$ & $23000$ & $18.8$ & $15.0$ & $14.0$ & $11.5$ & $14.1$ & $13.5$ & $23$ & $13.7$ & $13.8$\\
{\NV}$- \rm 9$ & $31$   & $0.25$ & $-1.6$ & $-3.6$ & $54$ & $22000$ & $19.6$ & $15.8$ & $14.8$ & $12.4$ & $15.0$ & $14.3$ & $53$ & $50.0$ & $50$\\

\hline
\enddata
\vglue -0.05in

\tablecomments{
\baselineskip=0.7\baselineskip
Column densities are listed in logarithmic units. In the
intermediate--ionization phase, both the photoionization and
the collisional ionization cases are listed.}

\label{tab:tab4}
\end{deluxetable}

\begin{deluxetable}{lccccccccccccccc}
\tablenum{5}

\tabletypesize{\scriptsize}
\rotate
\tablewidth{0pt}
\tablecaption{System C at $z=0.9343$}
\tablehead{
\colhead{} &
\colhead{$v$}  &
\colhead{$Z$} &
\colhead{$\log U$} &
\colhead{$\log n_H$} &
\colhead{size} &
\colhead{$T$} &
\colhead{$N_{\rm tot}({\rm H})$} &
\colhead{$N({\HI})$} &
\colhead{$N({\MgII})$} &
\colhead{$N({\SiIV})$} &
\colhead{$N({\CIV})$} &
\colhead{$N({\NV})$} &
\colhead{$b({\rm H})$} &
\colhead{$b({\rm Mg})$} &
\colhead{$b({\rm N})$} \\
\colhead{} &
\colhead{[{\kms}]} &
\colhead{[$Z_{\odot}$]} &
\colhead{} &
\colhead{[{\cc}]} &
\colhead{[kpc]} &
\colhead{[K]} &
\colhead{[{\cmsq}]} &
\colhead{[{\cmsq}]} &
\colhead{[{\cmsq}]} &
\colhead{[{\cmsq}]} &
\colhead{[{\cmsq}]} &
\colhead{[{\cmsq}]} &
\colhead{[{\kms}]} &
\colhead{[{\kms}]} &
\colhead{[{\kms}]}
}

\startdata
\multicolumn{16}{c}{\sc {\MgII}~Phase}\\
\hline
{\MgII}$- \rm 12$ & $1042$ & $1.25$ & $-2.0$ & $-3.2$ & $4$ & $10000$ & $18.9$ & $15.7$ & $12.1$ & $14.1$ & $15.0$ & $13.4$ & $20$ & $7.5$ & $8.2$\\

\hline
\multicolumn{16}{c}{\sc Offset~Phase} \\
\hline
{\CIV}$- \rm off$ & $1005$  & $0.063$ & $-1.6$ & $-3.6$ & $6$ & $30000$ & $16.6$ & $14.8$ & $8.4$ & $10.5$ & $13.3$ & $12.7$ & $26.3$ & $14.8$ & $15.3$\\
\hline\hline

&
$v$ &
$Z$ &
$\log U$ &
$\log n_H$ &
size &
$T$ &
$N_{\rm tot}({\rm H})$ &
$N({\HI})$ &
$N({\OVI})$ &
$N({\SiIV})$ &
$N({\CIV})$ &
$N({\NV})$ &
$b({\rm H})$ &
$b({\rm O})$ &
$b({\rm N})$ \\
 &
[{\kms}] &
[$Z_{\odot}$] &
 &
[{\cc}] &
[kpc] &
[K] &
[{\cmsq}] &
[{\cmsq}] &
[{\cmsq}] &
[{\cmsq}] &
[{\cmsq}] &
[{\cmsq}] &
[{\kms}] &
[{\kms}] &
[{\kms}] \\

\hline
\multicolumn{16}{c}{\sc {Diffuse}~Phase} \\
\hline
{\OVI}$- \rm ph$ & $1042$ & $1.0$ & $-0.5$ & $-4.7$ & $13$ & $31000$ & $17.9$ & $12.9$ & $14.0$ & $7.6$ & $6.1$ & $12.6$ & $45.6$ & $40.0$ & $40.1$\\
\hline
{\OVI}$- \rm co$ & $1042$ & $1.0$ & \nodata & \nodata & \nodata & $320000$ & $17.9$ & $12.0$ & $14.0$ & $10.1$ & $11.7$ & $12.1$ & $80.6$ & $40.0$ & $
40.6$\\

\hline
\enddata
\vglue -0.05in

\tablecomments{
\baselineskip=0.7\baselineskip
Column densities are listed in logarithmic units. For cloud {\MgII}$- \rm 12$, the abundance
of silicon was increased by $0.4$~dex, and the abundance of nitrogen was decreased by
$0.4$~dex.  For the high--ionization phase, both the photoionization and the collisional
ionization cases are listed.}

\label{tab:tab5}
\end{deluxetable}

\end{document}